\documentclass[manuscript,authorversion,nonacm]{acmart}
\usepackage{amsmath,amsfonts}
\usepackage{algorithmic}
\usepackage{graphicx}
\usepackage{textcomp}
\usepackage{xcolor}
\usepackage{ulem}
\usepackage{framed}
\usepackage{multirow}
\usepackage{orcidlink}
\usepackage{subcaption} 
\usepackage{booktabs}
\usepackage{svg}
\usepackage{soul}
 \usepackage{pythonhighlight}
 \usepackage{lipsum} 
 \usepackage{lscape} 
\usepackage[most]{tcolorbox} 
\usepackage{adjustbox} 
\usepackage[T1]{fontenc}  

\newtcolorbox{summarybox}{
    colback=black!20, 
    colframe=black, 
    coltext=black, 
    boxrule=0.5mm, 
    arc=3mm, 
    boxsep=3mm, 
    left=3mm, 
    right=3mm, 
}
\newcommand{\JIEW}[1]{\textcolor{black}{#1}}
\newcommand{\FHF}[1]{\textcolor{black}{#1}}
\newcommand{\JIE}[1]{\textcolor{black}{#1}}
\newcommand{\JW}[1]{\textcolor{black}{#1}}
\newcommand{\JWR}[1]{\textcolor{black}{#1}}
\newcommand{\JWRR}[1]{\textcolor{black}{#1}}
\newcommand{\NA}{---}

\AtBeginDocument{%
  \providecommand\BibTeX{{%
    \normalfont B\kern-0.5em{\scshape i\kern-0.25em b}\kern-0.8em\TeX}}}

\acmJournal{TOSEM}
\acmVolume{}
\acmNumber{1}
\acmArticle{}
\setcopyright{acmlicensed}
\acmISBN{1557-7392} 
\acmYear{2025}
\acmMonth{1}
\acmDOI{10.1145/3715109}

\thanks{This article is published in ACM Transactions on Software Engineering and Methodology (TOSEM)}

\begin{document}


\title{\JW{HumanEvalComm: Benchmarking the Communication Competence of Code Generation for LLMs and LLM Agent}}

\author{Jie JW Wu\orcidlink{0000-0002-7895-2023}}
\affiliation{%
  \institution{University of British Columbia, Kelowna}
  \streetaddress{3333 University Way}
  \city{Kelowna}
  \state{B.C.}
  \postcode{V1V 1V7}
  \country{Canada}
}
\email{jie.jw.wu@ubc.ca}

\author{Fatemeh H. Fard}
\affiliation{%
  \institution{University of British Columbia, Kelowna}
  \streetaddress{3333 University Way}
  \city{Kelowna}
  \state{B.C.}
  \postcode{V1V 1V7}
  \country{Canada}
}
\email{fatemeh.fard@ubc.ca}


\begin{abstract}

Large language models (LLMs) have significantly improved their ability to perform tasks in the field of code generation. However, there is still a gap between LLMs being capable coders and being top-tier software engineers. 
The most recent trend is using LLM-based agents to iterate the code generation process. 
Based on the observation that top-level software engineers often ask clarifying questions to reduce \textit{Ambiguity} in both requirements and coding solutions, we argue that the same should be applied to LLMs for code generation tasks. 
For this purpose, we define the communication skills of LLMs as ``being able to ask clarifying questions when the description of the code generation problem has issues''. In this study, we restrict these issues to three matters from the software requirement engineering field: inconsistent requirements, ambiguous requirements, and incomplete requirements. 
By asking probing questions about the requirements of problem descriptions before generating the final code, the challenges of programming with LLMs, such as unclear intent specification may be alleviated, resulting to a correct code in the initial iterations. 

In this work, we conducted an empirical study on the benchmark and analysis of the communication skills of LLMs for code generation. 
We created a new benchmark, HumanEvalComm, by modifying problem descriptions according to three issues mentioned above, \textit{Inconsistency}, \textit{Ambiguity}, \textit{Incompleteness}. We then experimented on HumanEvalComm with different Code LLMs, and a new LLM agent approach, \textit{C\underline{o}de \underline{C}l\underline{a}rificatio\underline{n} \underline{a}nd \underline{G}eneration \underline{A}ge\underline{n}t} (Okanagan), to identify and ask questions in ambiguous parts from code and descriptions for further refining the generated code. In the evaluation, we introduced an \textit{LLM-based evaluator} and created \textit{Communication Rate }and \textit{Good Question Rate} as the evaluation metrics to represent the ratio of questions asked and questions with good quality in responses. We found that more than 60\% of responses from Code LLMs still generate code rather than ask questions
when the problem descriptions are manually modified according to different clarification categories.
The Pass@1 and Test Pass Rate of most Code LLMs drop by 35\% $\sim$ 52\% and by 17\% $\sim$ 35\% respectively, with
statistical significance in each category for over 75\% numbers. Okanagan, as an LLM agent approach that uses LLM such as ChatGPT 3.5, effectively increases the Communication Rate and Good Question Rate by an absolute 58\% and 38\%, respectively. Thus, Okanagan boosts Pass@1 and Test Pass Rate by an absolute 8\% and 7\%, respectively, when the problem descriptions are modified based on given clarification categories. This result indicates the potential for achieving more effective communication capability using LLM agent.  Our benchmark and full code are publicly available at \href{https://github.com/jie-jw-wu/human-eval-comm}{https://github.com/jie-jw-wu/human-eval-comm}.

\end{abstract}





\maketitle

\begin{quote}
``Asking a good question can be valuable in and of itself, irrespective of the answer. It communicates your respect for the other person.'' \phantom{||} - Adapted from the Iowa Peace Institute Message\\
\end{quote}

\section{Introduction}

Large language models (LLMs) \cite{vaswani2017, svyatkovskiy2020, wang2021, feng2020}, such as OpenAI's Codex~\cite{chen2021evaluating}, AlphaCode~\cite{li2022competition}, and CodeGen~\cite{nijkamp2022codegen}, possess a significantly capable ability to generate code snippets from natural language requirements. However, there are several reported issues in LLMs, including problems with intent specification, problem decomposition~\cite{sarkar2022like}, code quality, and overconfidence~\cite{liu2023no,liu2023refining}, as well as usability~\cite{liang2023understanding}. These issues indicate that there is still a substantial gap between using LLM as a seasoned coder \cite{rabinovich2017, ye2020, alon2019, bui2021, tufano2020} and using LLM as a software engineer. As the responsibility of software developers encompasses more than just writing code, current LLMs cannot fully replace professional software developers~\cite{sarkar2022like,borji2023categorical}. 
At a high level, the gap lies in several critical aspects of software development beyond coding, such as effective communications, \JWR{explanation}, requirements, design, domain knowledge, and the broader context of relevant projects and components \cite{nguyen2022, sobania2022, vaithilingam2022, siddiq2022,palacio2024toward}. 
\JWR{LLM-based agents are different from LLMs in that LLMs are underlying models for text or code generation with pre-trained capability, while LLM-based agents deploy these models in a more structured, action-driven environments~\cite{guo2024large}. }
\FHF{Although some LLM-based agent systems have got a lot of attention, \JWR{such as SWE-Agent~\cite{yang2024swe}, CoderAgent~\cite{zhang2024codeagent}, AutoGPT~\cite{auto-gpt} and many others ~\cite{yao2023react,cognitionlabs,shinn2023reflexion,schick2023toolformer}}, there is no study that investigates the reported issues and systematically integrates them with the LLM agent approach for code generation.}
In this paper, we are interested in applying the communication lens to inspect the gap, given that we envision effective communication as a critical capability that ensures the necessary information is obtained for completing the coding tasks. 

Let us take a step back to compare the communications of LLMs and software developers. The current LLMs are typically evaluated by generating code in one or multiple attempts from one-off problem descriptions, without further conversational inputs~\cite{chen2021evaluating,austin2021program,li2022competition}. 
This means when the input problem description is error-prone or incomplete without full context, the model has to generate the code without the chance to clarify questions that are necessary to ensure the correctness of the code. \JIE{In the literature, the communication capability (defined below) of Code LLM~\cite{fan2023large,zan2023large} and LLM agent~\cite{rasheed2024codepori,xi2023rise} is underrepresented and thus rarely emphasized and evaluated in the field of code generation.} On the contrary, given a software engineering task in real-world enterprises, professional developers use various ways of communication, such as asking more questions in 1:1 conversations, group meetings, and Slack channels to obtain more information and reduce \textit{Ambiguity} about the detailed requirements, the context of the projects, and the design alternatives. Proactive and effective communication is a critical skill in practice for top-level software developers to accomplish their software engineering tasks reliably with high quality~\cite{whitehead2007collaboration,pressman2005software,mistrik2010collaborative,mcchesney2004communication,jazayeri2004education}.

Inspired by this behavior, our motivation in this work is to study and evaluate the potential of LLMs on code generation from the dimension of effective communication skills. We argue that the evaluation of the communication capability of Code LLMs is, although underrepresented in literature, essential for the long-term success of AI systems in completing the coding and software engineering tasks~\cite{hassan2024rethinking}. Thus, we intend to fill this literature gap in this research \FHF{for the code generation task}. 
For highly specialized task of code generation, we argue that the AI system should proactively recognize which information is missing, and find these missing pieces to be able to complete the task with high quality and rigorousness, instead of just executing the given task and generating low-quality code as a result. Formally, the \textit{communication capability}, also referred to as \textit{communication skills} or \textit{communication competency}, in this study is defined as follows: \textbf{when the requirements are incomplete, inconsistent, or ambiguous in a programming problem, and the model is prompted to either generate code or ask clarifying questions, how good the model is in asking clarifying questions to recover the requirements necessary for solving the problem correctly}. \JIE{We use the terms ``LLMs'' and ``Code LLMs'' interchangeably to represent LLMs for code generation tasks in this paper. }



\JIE{In this research, we conducted the first systematic empirical study on the communication skills of LLMs in code generation tasks. First, we created a benchmark dataset, \textbf{HumanEvalComm}, for evaluating the degree of communication skills when generating code, based on the widely studied HumanEval code generation benchmark~\cite{chen2021evaluating}. We constructed the benchmark by manually modifying the requirements in the original problem description based on concepts in Requirement Engineering (RE)~\cite{tukur2021requirement,dermeval2016applications}. To achieve this, we created a taxonomy of clarification types: \textit{\textit{Ambiguity}}, \textit{\textit{Inconsistency}}, and \textit{\textit{Incompleteness}} (See Section~\ref{sec2}). Based on the taxonomy, we then changed each problem description by applying one or a combination of clarification types. Based on the new HumanEvalComm benchmark, we further evaluated different models to inspect the degree of their communication skills when certain information is manually modified to be \textit{ambiguous}, \textit{inconsistent}, or \textit{incomplete} in the problem description. In the evaluation, we introduced an LLM-based
evaluator and proposed new evaluation metrics to effectively measure the communication skills of the models.
We also proposed a LLM agent approach, \textit{C\underline{o}de \underline{C}l\underline{a}rificatio\underline{n} \underline{a}nd \underline{G}eneration \underline{A}ge\underline{n}t} (\textbf{Okanagan}), as an LLM-based agent with multi-round structure and customized prompt for code generation task. A key feature of Okanagan is the ability to ask clarifying questions about the input problem descriptions needed for generating correct code.}

\JIE{In terms of findings, for manual modifications \FHF{using HumanEvalComm}, more than 60\% of responses from Code LLMs still generate code. Typically, the Pass@1 and Test Pass Rate of Code LLMs drop by 35\% $\sim$ 52\% and by 17\%
$\sim$ 35\%, respectively. Among the three clarification types, the \textit{Incompleteness} category results in higher communication
rates and Good Question Rates, but lower Pass@1 and Test Pass Rate than the \textit{Ambiguity} and \textit{Inconsistency}
categories for Code LLMs. \textit{Okaganan}, the proposed LLM agent approach that uses ChatGPT 3.5 as LLM,   effectively increased Communication Rate and
Good Question Rate by an absolute 59\% and 5\%, respectively. This resulted in an increase in Test Pass Rate and Pass@1 by 25\% and 15\%, respectively. This result indicates the potential for more effective communication capability for LLM agent compared with Code LLMs.}

To summarize, we have made the following contributions:

\begin{itemize}
 
    \item We created a new benchmark, \textbf{HumanEvalComm}, for evaluating the degree of communication skills of LLMs for code by manually modifying the requirements in the original problem description based on RE concepts: clarification types of \textit{\textit{Ambiguity}}, \textit{\textit{Inconsistency}}, \textit{\textit{Incompleteness}}.
    
    \item We proposed an LLM-agent approach, \textit{C\underline{o}de \underline{C}l\underline{a}rificatio\underline{n} \underline{a}nd \underline{G}eneration \underline{A}ge\underline{n}t} (\textbf{Okanagan}), to \FHF{enhance the communication capability of the models, and thus lead to better code generation capability, in terms of Pass@1, based on asking clarifying questions when the problem description is ambiguous, inconsistent, or incomplete}. \JIE{The contribution of Okanagan is a multi-round structure with customized prompts for asking clarifying questions when needed in code generation tasks. } 

    \item We conducted the first empirical study on the evaluation of communication competence in code generation task for both Code LLMs and Okanagan on HumanEvalComm. \JIE{In the evaluation, we introduced \textit{LLM-based evaluator} and proposed two new evaluation metrics, \textit{Communication Rate} and \textit{Good Question Rate}, to effectively measure communication skills of the models.} 

\end{itemize}

Our benchmark and replication package are made public at \href{https://github.com/jie-jw-wu/human-eval-comm}{https://github.com/jie-jw-wu/human-eval-comm} to support open data and open science principles. 

The rest of the paper is structured as follows. Section~\ref{sec2} describes the benchmark construction of our research. Section~\ref{sec3} explains the design of our empirical study. Section~\ref{sec4} summarizes the results for RQs. Section~\ref{sec6} includes more analysis and discussions on the results. Threats to validity are explained in Section~\ref{sec7}, followed by summarizing the related works in Section~\ref{sec8}. Finally, Section~\ref{sec9} concludes this work. 


\section{Benchmark Construction}
\label{sec2}

\subsection{Benchmark Collection}
\textbf{Existing Benchmarks.} We first start by examining the existing benchmarks for code generation. To the best of our knowledge, all of the existing benchmarks (e.g., HumanEval~\cite{chen2021evaluating}, CoNaLa~\cite{yin2018learning}, Galeras~\cite{rodriguez2023benchmarking}, APPS~\cite{hendrycks2021measuring}, and recent SWE-bench~\cite{jimenez2024swebench}) in code generation are tasked with letting the model generate the code directly as prediction, without giving the model the opportunity to \textit{ask for} additional information. 
Notably, the input of these datasets is well-written and organized by professional human annotations. However, in real-world scenarios, the problem descriptions from humans could be a lack of computational thinking, unclear in the intent specification, or ambiguous in requirements~\cite{liang2023understanding}.

\textbf{HumanEvalComm Overview.} To assess the communication ability of Code LLMs and LLM-based agent, we chose to hand-craft a new benchmark based on a widely used code generation dataset, HumanEval~\cite{chen2021evaluating}. \JIE{Our objective is to modify the problem description based on RE concepts so that it should trigger clarifying questions, which are necessary for generating the correct code.} HumanEval is composed of 164 hand-crafted coding problems in Python and was created to evaluate the coding capabilities of Codex. Each problem has a function signature, docstring, body, and unit tests. The average number of ground-truth test cases per problem is 7.77. HumanEval is chosen as it is a benchmark dataset with test cases and is widely used for evaluating LLMs~\cite{ouyang2023llm,min2023beyond,rasheed2024codepori,zan2023large}. Using HumanEval, we changed each problem description manually to develop HumanEvalComm, which we will use for evaluation in our work. This is done using a taxonomy of clarification types as described below. \JWR{Besides, we provided explanation about why we chose manual method instead of automated method to modify problems in Section~\ref{sec:auto_modification}.}

\textbf{Taxonomy of Clarification Types.} \JIE{To modify the problem description in an organized way, we propose the following clarification types based on both the literature in Requirement Engineering (RE)~\cite{tukur2021requirement,dermeval2016applications} and our understanding of how feasible can the RE concepts be applied to problems in HumanEval:}

\begin{itemize}
    \item \textbf{\textit{Ambiguity}}: \JW{Some statements in the problem descriptions could be ambiguous and correspond to different concepts.}
    \item \textbf{\textit{Inconsistency}}: Some statements in the problem descriptions show conflict or inconsistency between each other.
    \item \textbf{\textit{Incompleteness}}: Some concepts or conditions are missing in the problem descriptions.
\end{itemize}

\begin{table*}[]
\begin{center} 
\begin{tabular}{|c|c|c|c|c|}
\toprule
Clarification Category  & \textit{Ambiguity} & \textit{Inconsistency} & \textit{Incompleteness} & \JWR{\textit{Count}} \\  
\midrule
 1a &\checkmark & & &\JWR{164}\\ 
\midrule
 1c & &\checkmark  & &\JWR{164}\\ 
\midrule
 1p & & &\checkmark  &\JWR{164}\\ 
\midrule
 2ac &\checkmark  &\checkmark  & &\JWR{162}\\ 
\midrule
 2cp & &\checkmark  &\checkmark & \JWR{34} \\
\midrule
 2ap &\checkmark  & &\checkmark &\JWR{74} \\ 
\midrule
 \JWR{Total} &\NA  &\NA &\NA &\JWR{762} \\ 
\bottomrule
\end{tabular}
\end{center}
\caption{\JWR{Problem descriptions and statistics with different combinations of clarification types being applied in HumanEvalComm. The smaller size for 2ac (same applies for 2cp and 2ap) is because we directly applied a combination of two clarification types from 1a, 1c strictly, and we create a new modified problem as 2ac only if applying a combination of 1a and 1c leads to a new problem description that is different from either 1a and 1c. 2cp and 2ap have smaller count because the ambiguous (a) or inconsistent (c) parts are removed in (p) for a large number of problems.}}
\label{Tab:clarification_types}
\end{table*}

\begin{table*}
\centering
\begin{center} 
\begin{tabular}{p{2cm}p{12cm}}
\toprule
Clarification Type  	& Problem Description \\
\midrule
N/A (Original) &  
\begin{python} 
def incr_list(l: list):
    """Return list with elements incremented by 1.
    >>> incr_list([1, 2, 3])
    [2, 3, 4]
    >>> incr_list([5, 3, 5, 2, 3, 3, 9, 0, 123])
    [6, 4, 6, 3, 4, 4, 10, 1, 124]
    """

\end{python} \\

\textit{Ambiguity} &
\JWR{The docstring is ambiguous as the number is not specified. }
\begin{python}

def incr_list(l: list):
    """Return list with elements incremented by a number.
    >>> incr_list([1, 2, 3])
    [2, 3, 4]
    >>> incr_list([5, 3, 5, 2, 3, 3, 9, 0, 123])
    [6, 4, 6, 3, 4, 4, 10, 1, 124]
    """
\end{python} \\
\textit{Incompleteness} &
\JWR{The docstring is an unfinished sentence. }
\begin{python}

def incr_list(l: list): 
    """Return list with elements incremented
    """
\end{python} \\

\textit{Inconsistency} &
    \JWR{Result of the two test examples conflict with the docstring description as well as each other}
\begin{python}
def incr_list(l: list):
    """Return list with elements incremented by 1.
    >>> incr_list([1, 2, 3])
    [3, 4, 5]
    >>> incr_list([5, 3, 5, 2, 3, 3, 9, 0, 123])
    [7, 5, 7, 4, 5, 5, 11, 2, 125]
    """
\end{python} \\
\textit{Inconsistency} \& \textit{Ambiguity} &
\JWR{The docstring is ambiguous with the unspecified number for incrementing, and the results of test examples conflict with the docstring.}
\begin{python}
def incr_list(l: list):
    """Return list with elements incremented by a number.
    >>> incr_list([1, 2, 3])
    [3, 4, 5]
    >>> incr_list([5, 3, 5, 2, 3, 3, 9, 0, 123])
    [7, 5, 7, 4, 5, 5, 11, 2, 125]
    """
\end{python} \\
\bottomrule
\end{tabular}
\end{center}
\caption{Example of HumanEvalComm built upon HumanEval. The modified problem descriptions are shown in this table for problem number 42 of HumanEval. Specifically, the descriptions of the problem were modified to be inconsistent, ambiguous, or incomplete. The main goal of the HumanEvalComm dataset is to evaluate the degree of communication.}
\label{Tab:dataset2_example}
\end{table*}

\JWR{\textbf{Guideline for Modifying Problem Description.}} For each problem description, 
we manually change the problem description with regard to different clarification types. \JWR{As different engineers might approach the same problem differently, therefore impacting the consistency and standardization of the modified problems~\cite{sim2003using}, we describe the guidelines for the manual work. } Modifying the problem descriptions is done manually by a software engineer with nearly a decade of experience in the industry. A second software engineer with more than 15 years of development experience reviewed the changed descriptions.
The disagreements were marked and discussed among the two annotators until they reached an agreement about the changes, according to the definitions of \textit{Ambiguity}, \textit{Inconsistency}, and \textit{Incompleteness} from RE.
Each problem description was read carefully, and modifications were applied to the problem description. 
The definitions and examples of ambiguous, inconsistent, or incomplete requirements were reviewed by both people before conducting the manual modification of the problem descriptions, although both of them have requirements engineering expertise. 

For each problem, we applied six different modifications: (1a) making the problem description ambiguous; (1c) modifying the description to be inconsistent; (1p) changing the problem description to make it incomplete. The next three modifications that we refer to as (2ac), (2cp), and (2ap) are a combination of the initial changes, being ``ambiguous and inconsistent'', ``inconsistent and incomplete'', and ``ambiguous and incomplete'', respectively. For any of the above modifications, our standard is that \JWR{\textbf{applying the modification to the problem should trigger clarifying questions, which are necessary for generating the correct code.}} \JWR{To ensure the standardization of the manual work, for each clarification category, the criteria to modify the problem and disagreements with the modified problems were thoroughly discussed in several discussion sessions and rounds to change each record. Besides, we have also added this to the Threats of Validity section in the paper. Below are the detailed guideline for each clarification category:} 
\begin{itemize}
    \item \textbf{\textit{\JWR{Clarification Category 1a}}}: Specifically, for 1a, to make the descriptions ambiguous, we tried to change the statement such that a human reader can interpret the statement in different ways. However, in practice, we found it very difficult to perform ambiguous modifications: adding \textit{Ambiguity} in description only may not be enough to trigger clarifying questions, since we have additional information such as test examples, common sense reasoning, and function signatures. In other words, the description becomes ambiguous, but the test examples and function signatures are also given in the description, so that the right requirements can be inferred and thus correct code can be generated, without having to ask clarifying questions. Our solution to this issue is that we apply both \textit{Ambiguity} and incorrectness to the description because it's much easier and safer to trigger clarifying questions using \textit{Ambiguity} and a certain level of incorrectness, instead of \textit{Ambiguity} only.  For example, changing the description from ``sort the array descendingly'' to ``sort the array'' may not trigger a question, because the function signature or test cases can imply the sorting is in descending order. However, changing the description to ``sort the array (descendingly or ascendingly)'' can trigger questions.
    \item \textbf{\textit{\JWR{Clarification Category 1c}}}: For 1c, to make the descriptions inconsistent, we mainly changed the examples such that the output of the example does not match or contradict the problem description. It should be noted that most of the problem descriptions in the HumanEval benchmark contain examples of test cases with the input and output. When applying inconsistent modification, for each problem, we changed the output of the test examples in a meaningful way rather than randomly, to enhance the contradiction between test examples and text description.  \JWR{Note that 1c is different from 1a, as 1c focuses mainly on changing the test examples so that they contradict the problem description, while 1a changes the requirements so that the correct requirement is combined with incorrect requirements, making it unclear and confusing which one is the correct requirement.}
    \item \textbf{\textit{\JWR{Clarification Category 1p}}}: In terms of 1p, we removed some parts of the description as incomplete modification. We made sure that after applying the incomplete modification, it's not possible to generate the correct code, without asking questions to recover the missing content.
    \item \textbf{\textit{\JWR{Clarification Category 2ac, 2cp, 2ap}}}: For 2ac, 2cp, and 2ap, we directly applied a combination of two clarification types from 1a, 1c, and 1p. For these cases, we create a new modification only if applying a combination of two types leads to a new description that is different from any of the two types. Therefore, for each problem, 1a, 1c, and 1p always exist, but 2ac, 2cp, or 2ap may not exist. Overall, the process of changing the descriptions took approximately 100 hours for initial modification and 30 hours to review and discuss the disagreements and come to a consensus.

\end{itemize}

Table~\ref{Tab:dataset2_example} shows an example of the original problem description and three modified versions for problem number 42 in HumanEval. 
\FHF{In this example, for \textit{Ambiguity}, ``incremented by 1'' is modified to ``incremented by a number'', forming an ambiguous description. For \textit{Incompleteness}, a part of the text description and example test cases are removed. For \textit{Inconsistency}, the output of examples is modified so that it contradicts the text description. For \textit{Inconsistency} and \textit{Ambiguity}, a combination of \textit{Inconsistency} and \textit{Ambiguity} is applied, making it a more challenging case to generate the correct code.}
It is worth mentioning that before constructing HumanEvalComm, we manually verified the problem descriptions in the HumanEval dataset and verified that the original problem descriptions do not have clarification issues (\textit{Ambiguity}, \textit{Inconsistency}, or \textit{\textit{Incompleteness}}), so we chose all of the 164 problems in HumanEval dataset in our evaluation.

\subsection{Evaluation Measurement}
\JIE{We introduce the following metrics to effectively evaluate the communication competency of the models in code generation tasks.}

\textbf{Communication Rate.} We propose the communication rate to evaluate the degree of communication skills for a given model. The communication rate is intended to capture the percentage of responses with clarifying questions instead of code for problems in HumanEvalComm. In the experiment, the prompt we use lets the model ``either generate Python3 code (Respond directly with code only with markdown), or ask clarifying questions''. Therefore, in this work, we define the communication rate as the percentage of responses with no code snippets (non-code) for the initial modified problem descriptions: 

\begin{equation*}
communicaton\_rate = \dfrac{\# initial\: model\: responses\: without\: code}{\# initial\: model\: responses}
\end{equation*}
In the experiment, we found that this simple metric that distinguishes whether the model returns code or non-code is already an effective approximation of communication skills.

\textbf{Good Question Rate.} In this research, we leverage a new \textit{LLM-based evaluator} 
to give a question quality label for clarifying questions returned by the models. The labels are \textit{Good} (The model asks insightful questions that help recover all the missing info), \textit{Fair} (The model asks OK questions, but the questions do not fully cover the missing info), \textit{Bad} (The model asks no questions or irrelevant questions that do not help at all \FHF{to recover the missing/clarifying information}). Given the question quality label, we define Good Question Rate as the percentage of responses with \textit{Good} question quality labels:

\begin{equation*}
\JIEW{good\_question\_rate = \dfrac{\# initial\: model\: responses\: with\:Good\: labels}{\# initial\: model\: responses}}
\end{equation*}



\textbf{Pass@K.} In evaluation, pass@k is a popular and widely used metric for evaluating the task of code generation~\cite{chen2021evaluating,min2023beyond,zhang2023unifying}. Pass@k is defined as the ratio of ``solved'' problems, in which as a problem is \JWR{`solved'} if any of the $k$ code samples pass all the tests. Hence, we used Pass@1 in our evaluation.

\textbf{Test Pass Rate.} Besides the widely used pass@k, \textit{Test Pass Rate} is also commonly used for evaluating code generation~\cite{ouyang2023llm,hendrycks2021measuring}. Specifically, the Test Pass Rate is defined as the proportion of successfully passed test cases in relation to the total number of test cases for LLM-generated code. This metric is useful in this work since it helps capture whether getting the right information by asking clarifying questions can indeed increase the correctness of generated code. 

\section{Empirical Study}
\label{sec3}
\subsection{Research Questions}
In this section, we describe the research questions that we explore in this study.



\textbf{RQ1: How do \FHF{Code} LLMs perform in communication competency when requirements in the problem descriptions are incomplete, inconsistent, ambiguous?}\\
\JW{The rationale of RQ1 is centered around understanding and examining the current Code LLMs regarding their communication capabilities in code generation. The aim is to provide an initial understanding of the limitations and areas where Code LLMs may fall short in their communication skills. We evaluated different Code LLMs on carefully curated problems in the new benchmark, HumanEvalComm, where problem descriptions are manually modified to be incomplete, inconsistent, and ambiguous. We evaluated and compared the results of Code LLMs for different clarification categories, where one or two clarification types are applied to the original problems.}


\textbf{RQ2: How \FHF{does Okanagan} perform compared with Code LLMs in terms of communication skills?}\\
Given the recent advances in LLM-based agent in addressing various applications~\cite{xi2023rise}, RQ2 aims to investigate the communication capabilities of our LLM agent approach, \textit{Okanagan}, in comparison with Code LLMs. Therefore, we evaluated Okanagan \JIE{which has a multi-round structure with customized prompts for code generation tasks.} We analyzed and compared the results of Okanagan and Code LLMs.


\textbf{\JWR{RQ3: How to assess the LLM-based evaluator and the evaluation metrics generated by the evaluator?}}\\
\JWR{The aim of RQ3 is to investigate and assess the LLM-based evaluator used in our evaluation. In our evaluation, we used the \textit{LLM-based evaluator} to 1) rate the quality of the clarifying questions and 2) generate an answer to reply to the clarifying questions from the models~\cite{kotonya2023little,gao2024llm}. Therefore, it's important to evaluate the reliability of the LLM-based evaluator in our experiment. We conducted a manual evaluation of LLM-based evaluator where we recruited students at the University of British Columbia to manually assess the results of LLM-based evaluator and manually mark the quality of models' responses.}

\textbf{\JWR{RQ4: What are the impacts of different prompt strategies and hyperparameters?}}\\
\JWR{The goal of RQ4 is to investigate the impact of different prompt variants and hyperparameters important for the evaluation. We conducted experiments to evaluate the results of 1) using different hyperparameters, including temperatures and the ratio of removed content in the problem description for 1p, and 2) using different prompt strategies.} 

\subsection{Methodology Overview}
\begin{figure}[h]
  \centering
  \includegraphics[width=\textwidth]{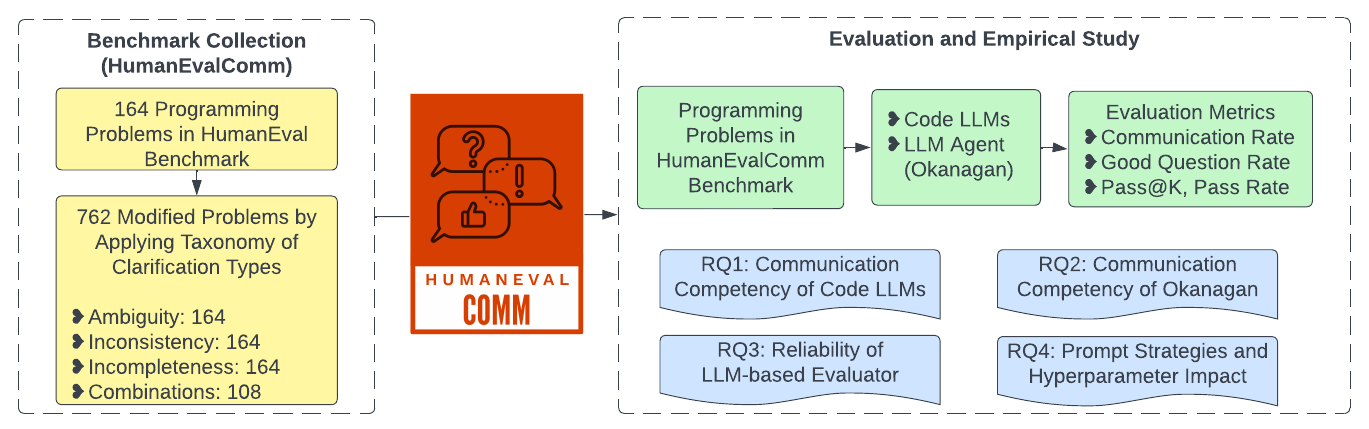}
  \caption{The visual illustration of the methodology \JWR{on HumanEvalComm benchmark (with statistics) and} the evaluation of communication skills for Code LLMs and LLM Agent. }
  \label{fig:method_overview}
\end{figure}

\textbf{Overview.} Figure~\ref{fig:method_overview} shows the overview of our methodology for collecting the benchmark and conducting the empirical study. 
We first create the HumanEvalComm benchmark, by modifying 164 problem descriptions of the original HumanEval benchmark for code generation tasks using the taxonomy of clarification types, as described in Section~\ref{sec2}. 
Second, we conduct evaluation and empirical studies based on HumanEvalComm to evaluate the communication competency of different models\JWR{, including Code LLMs and LLM Agents, as described in section~\ref{sec:codellms} and section~\ref{sec:ok}}.

\begin{figure}[h]
  \centering
  \includegraphics[width=\textwidth]{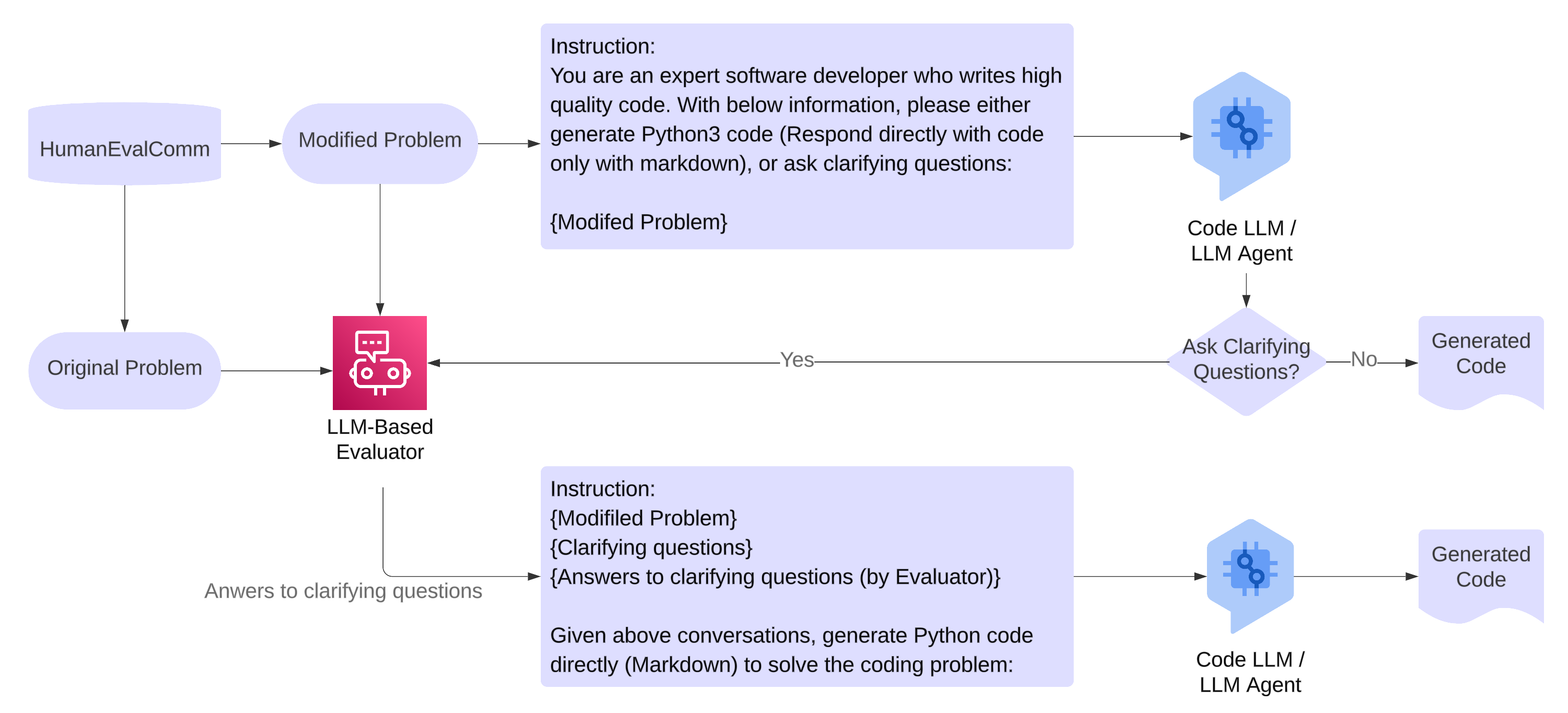}
  \caption{Flowchart for the evaluation of models, either Code LLMs or Okanagan (LLM agent), in communication capability.}
  \label{fig:eval_overview}
\end{figure}

Figure~\ref{fig:eval_overview} shows the flowchart for the evaluation of models, Code LLMs, and Okanagan. For each programming problem in the HumanEvalComm, there are up to six modified problem descriptions as described earlier in Table~\ref{Tab:clarification_types}. 
For each modified problem, a prompt is used as the input of the model to either generate code or ask clarifying questions if needed. Then, if the model asks clarifying questions rather than generates code directly, the questions are sent to an \textit{LLM-based Evaluator}, which evaluates the questions and generates a reply to answer the questions, based on all of the available information, including the modified problem, original problem, and the clarifying questions. Finally, the answers and the previous conversations are sent to the model to generate the code again directly. \JWR{We have also added examples in Section~\ref{sec:examples} to better understand the interaction between the evaluated models and the prompts. }

\JIE{\textbf{LLM-based evaluator}. With the advances of LLMs, a recent series of work has been proposed to use the powerful LLMs as the reference-free evaluators on Natural Language Generation (NLG) tasks~\cite{kocmi2023large,wang2023chatgpt,kim2023better,kotonya2023little,gao2024llm}. Given the expensive human efforts of human evaluations, we used the \textit{LLM-based evaluator} to generate an answer to reply to the list of clarifying questions from the models~\cite{kotonya2023little,gao2024llm}. We prompted the LLM-based evaluator with the modified problem, original problem, and clarifying questions. The role of the LLM-based evaluator in this work is to 1) generate answers to the clarifying questions, and 2) calculate \textit{Good Question Rate}, represented by an integer. The Good Question Rate is one of the evaluation metrics in our experiment. As for implementation, we used GPT 3.5 in the LLM-based evaluator in the evaluation. We tested both zero-shot and one-shot prompting, but in our evaluation, we found that one-shot prompting does not improve performance. This also aligns with the finding in literature~\cite{kotonya2023little,gao2024llm}.} The detailed prompt of LLM-based evaluator is shown as follows. 

\begin{framed}
\textcolor{black}{\textbf{Prompt for LLM-Based Evaluator:}}\\
The original description of a coding problem is modified so that the requirements become inconsistent, incomplete, or ambiguous. Given the modified description, some clarifying questions were raised to clarify the description. Given the original and modified problem description, evaluate the quality of the clarifying questions. Please provide an integer representing the quality of questions (3: Good questions that recover the modified requirements; 2: Fair questions but they cannot help recover the modified requirements; 1: No questions).\\  QUALITY=[your int] \\ Please also provide answers to the clarifying questions to recover the modified requirements in the original problem description compared to the modified one. If there are no clarifying questions at all, return empty answers. \\ ANSWERS=```[your answer]'''\\
Please strictly follow the format QUALITY=[the int] and ANSWERS=```[the answer]''' in the response! Surround your answer with markdown! \\ \\ \#\#\# Questions: \{clarifying\_questions\} \\ \#\#\# Modified Problem Description: \{problem\} \\ \#\#\# Original Description: \{original\_problem\} \\
\end{framed}

We manually checked the results of the LLM-based evaluator to see whether the output of the evaluator, the generated answers, and the evaluation of the questions, were correct. Overall, the generated answers and Good Question Rates are reasonable, but we do see some mistakes in both the generated answers and the Good Question Rates. For the Good Question Rates, there are some cases where the Good Question Rate from the LLM-based evaluator is 2 or 3 when there are no clarifying questions: For example, the ``questions'' from the model are a combination of explanation and code, but no clarifying questions. This is somewhat related to the reported limitation~\cite{gao2024llm} that LLM evaluators prefer to give high scores to responses that conflict with the facts in the dialogue history~\cite{liu2023evaluate}.
For generated answers, we sometimes observe that the provided answers do not recover the original requirements, due to either the evaluator itself or ``no clarifying questions'' mentioned above. To mitigate this issue, we have optimized the prompt for the LLM-based evaluator several times and checked the results manually. This includes adding sentences like ``Please strictly follow the format QUALITY=[the int] and ANSWERS=[the answer] in the response!'', and ``Surround your answer with markdown!'' which eliminated many cases of format errors (answers and rates cannot be extracted correctly). Although the LLM-based evaluator shows effectiveness in our task, we understand that LLM-based NLG evaluation is still challenging~\cite{gao2024llm,liu2023evaluate}, and future work is required to address the errors mentioned above.


\subsection{Code Large Language Models}
\label{sec:codellms}


\JIE{We performed our evaluation on five widely used LLMs. This includes three open-sourced instruction-tuned Code LLMs, one open-sourced instruction-tuned LLM, and one commercial LLM. For open-source models, we used models with the largest possible model size within our limited computing resources in our evaluation.}

 \begin{itemize}
\item \JIE{\textbf{CodeLlama} (Instruction tuned version, 13B)~\cite{roziere2023code} is an open-source LLM released by Meta for coding, built on top of Llama 2, with foundation models and instruction models. CodeLlama was chosen because of its wide usage and top performance in HumanEval. We tested the instruction model CodeLlama-Instruct-13B in our experiment since we did not have the computing resources to run models with 34B. The same applies to the rest open-source models.}

\item \JIE{\textbf{DeepSeek Coder} (Instruction tuned version, 7B)~\cite{guo2024deepseek} is an open-source Code LLM trained on both 87\% code and 13\% natural language. Each of the models was pre-trained on 2 trillion tokens. We selected this model because it achieved top 5 performance in Big Code Models Leaderboard~\cite{bigcodemodelshf} on the HuggingFace platform. The Big Code Models Leaderboard~\cite{bigcodemodelshf} evaluates the performance of base multilingual code generation models on the HumanEval benchmark and MultiPL-E.  We used the model of 7 billion parameters in the evaluation. }

\item \JIE{\textbf{DeepSeek Chat} (Instruction tuned version, 7B)~\cite{bi2024deepseek} is an open-source LLM released by DeepSeek AI, trained on datasets of 2 trillion tokens. We selected this model because we wanted to evaluate the communication skills of models trained from different sources such as natural languages, code, and a combination of both. We compared its performance with the DeepSeek Coder to understand whether more natural languages in pre-training are beneficial to communication skills. We used the model of 7 billion parameters in the evaluation. }

\item \JIE{\textbf{CodeQwen1.5 Chat} (Instruction tuned version, 7B)~\cite{bai2023qwen} is an open-souce Code LLM released by Qwen Team, trained on 3 trillion tokens of code data. CodeQwen1.5 Chat is the Code-Specific version of Qwen1.5. The model is a transformer-based decoder-only language model and includes group query attention (GQA) for efficient inference. We selected this model because it achieved top 5 performance in Big Code Models Leaderboard~\cite{bigcodemodelshf}.}

\item \JIE{\textbf{ChatGPT}, released by OpenAI are powerful models for generation tasks. We used parameter-frozen versions of models (gpt-3.5-turbo-0125) to ensure the reproducibility of the evaluation results. }
\end{itemize}

\JIE{Note that all of the evaluated models above are instruction-tuned models because, in the evaluation, the ability to ask clarifying questions with the given prompts is needed for the models. Besides instruction-tuned models, there are also foundation models, but we didn't report results for foundation models. We found that foundation models without instruction tuned are not suitable for our evaluation, because their task is only to complete code and are not capable of instructions such as ``either generate code or ask clarifying questions''.}





\begin{figure}[h]
  \centering
  \includegraphics[width=\textwidth]{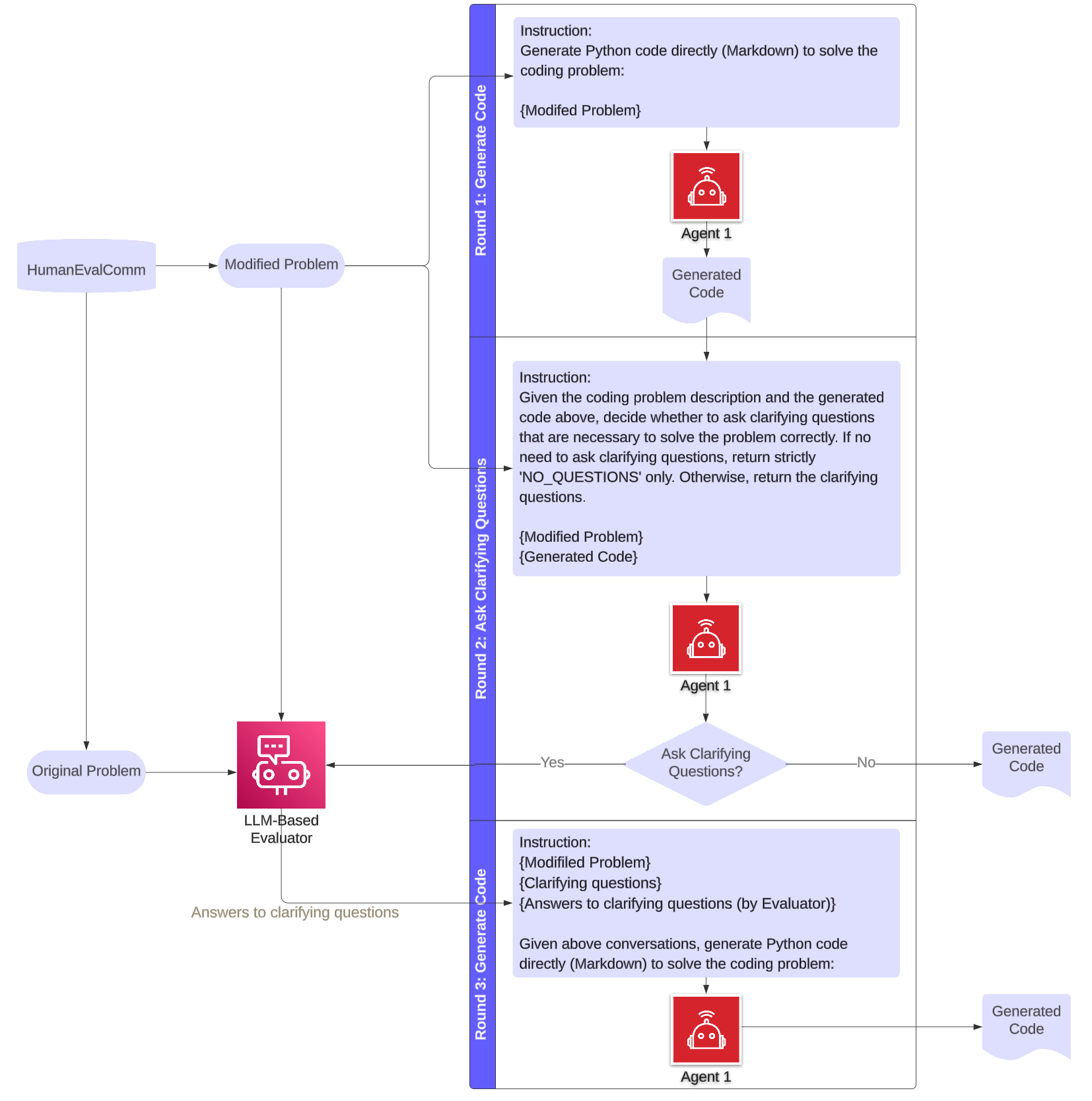}
  \caption{\JW{An illustration of the process of Okanagan, an LLM agent approach.} }
  \label{fig:okanagan_example}
\end{figure}

\subsection{LLM-Agent Approach (Okanagan)}
\label{sec:ok}

Following the recent works in LLM agent, including \textit{collaboration mechanisms for LLM agents}~\cite{zhang2023exploring} and \textit{self-correcting strategies of LLMs}~\cite{pan2023automatically}, we proposed and evaluated an LLM agent approach, \textit{Okanagan},  that leverages \JIE{multi-round structure and customized prompt format} for asking clarifying questions in code generation tasks. Figure~\ref{fig:okanagan_example} provides a visual illustration of the process of Okanagan. Specifically, we introduce three rounds in Okanagan:

 \begin{itemize}
\item Round 1: the agent generates code directly given the modified problem.
\item Round 2: the agent generates clarifying questions (if needed) given the modified problem and generated code. \JIE{If no questions, directly return the code generated in Round 1.}
\item Round 3: the agent generates code again, given the above conversation history (including the modified problem, clarifying questions, and their answers).
\end{itemize}

This structure is inspired by the existing LLM agents approach ~\cite{zhang2023exploring}, with three rounds and customized prompts for our task of code generation. \JIE{In terms of actions in each round, the action in Round 1 is to generate code. The action in Round 2 is to ask clarifying questions. As mentioned above, the code in Round 1 is returned if no questions are asked. Otherwise, a reflection is conducted to generate code again with the previous conversation history that includes clarifying questions and answers provided by the LLM-based evaluator.} We adopted this structure because it can be easily extended to different parameter values. For example, we stop at Round 3 (in other words, we set the total number of rounds to 3) in our evaluation, but we can set a different number of rounds in theory. Besides the number of rounds, other parameters can be changed as well. We describe the set of parameters for Okanagan as follows: 
\begin{enumerate}
    \item  \textit{number of agents} (default is 1).
    \item  \textit{number of rounds} (default is 3).
    \item \textit{Action in each round}: Generate code or ask questions \JIE{(default is: Round 1 - Generate code, Round 2 - Ask questions, Round 3 - Generate code with Reflection)}
    \item \textit{thinking pattern}: Debate or Reflection~\cite{zhang2023exploring}. (default is Reflection\footnote{\JIE{Debate can be used only when multiple agents are used. In the implementation of Okanagan, we use a single agent, and thus in Round 3, the single agent reflects on the generated code in Round 1 based on the additional information in Round 1 (generated code) and Round 2 (clarifying questions and their answers provided by LLM-based evaluator). }}) 
\end{enumerate}

\JIE{Note that we tried to minimize the complexity of Okanagan using default parameters such as one agent and three rounds}, but in future work, the structure can easily scale from single-agent to multi-agents by setting the parameter for the number of agents. If more than one agent is used, and when the thinking pattern is \textit{Debate} in a given round, the agents would exchange their previous responses as a way of collaboration. Given our specific task, compared with~\cite{zhang2023exploring}, we added a new parameter in Okanagan: action in each round, to indicate the action for agents in a given round. 

\JIE{\textbf{Okanagan Implementation in Evaluation}. In our evaluation, we tested Okanagan using the default parameters as mentioned. We used ChatGPT 3.5 as the LLM in each of the three rounds in Okanagan mainly for easier comparison with ChatGPT 3.5. For other LLM agent methods to compare in evaluation, we searched other publicly available LLM agent but did not find an appropriate open-sourced LLM agent implementation for code generation task with a focus and potential action to ask clarifying questions. } \JWR{
Table~\ref{tab:apx_prompts_comm} in Appendix shows the prompts used in Okanagan. Besides, we also switched base model in Okanagan to ensure a fair comparison. Concretely, we implemented the method using another open-sourced model (DeepSeek Coder) instead of ChatGPT 3.5 as the base LLM. }

\subsection{Experiment Setup}
\JWR{\textbf{Computational Resources.}} In experiments, we implemented our evaluation in Python 3.12. We partially used the code from ~\cite{ouyang2023llm} on the Non-Determinism of ChatGPT and from~\cite{min2023beyond} on testing open-source models. All of the experiments for ChatGPT and Okanagan were conducted on a server with an Intel i7-6700K CPU (4.00 GHz), 32 GB RAM. The other experiments for open-source models that require GPUs were conducted on an Intel Xeon Gold 6130 CPU (2.1GHz), 44 GB RAM, and 4 GPUs (Tesla V100-SXM2-16GB). \JIEW{The names of the HuggingFace models we use in the experiments are deepseek-coder-6.7b-instruct, deepseek-llm-7b-chat, CodeQwen1.5-7B-Chat, and CodeLlama-13b-Instruct-hf. }

\JWR{\textbf{Cost and Runtime.} Regarding the expenditure of the experiments, excluding AgentCoder, the total cost in using ChatGPT 3.5 is less than 15 USD. This cost includes the use of ChatGPT 3.5 from the LLM-based evaluator for all the models, and the usage in Okanagan and ChatGPT as models themselves. Separately, the cost for running AgentCoder alone in HumanEvalComm is about 36 USD, as it involves multiple OpenAI API calls within AgentCoder. The cost of ChatGPT 3.5 (`gpt-3.5-turbo-0125') is \$0.0005/1k input and \$0.0015/1k output. The total runtime for Okanagan or ChatGPT as the model in the experiment is less than half an hour to be conservative. The runtime for other open-source models depends on the computing resources such as memory and GPUs. Note that we have tried our best to reduce the redundant prompts between dialogues for simplcity and cost saving purpose: we used two prompts for a given model in each problem, which are necessary for a model to gather answers to its clarifying questions and generate correct code. For LLM-based evalator, we used only one prompt that generates both the answer to the question and the question quality label used in Good Question Rate. }

\JWR{\textbf{Statistical Test Configuration.} In this study, for each clarification category, we performed an independent two-sample t-test to assess whether there is a statistically significant difference between the correlations of original pass rates in HumanEval (specifically, Pass@K and Test Pass Rate) and the pass rates in HumanEvalComm. We assume the null hypothesis $H_0$ that there is no significant difference between Test Pass Rate or Pass@1 across the groups. The alternative hypothesis $H_1$ posits that there is a significant difference in the Test Pass Rate or Pass@1 between the groups. The test compares the means of two independent groups, testing the null hypothesis that the means are equal. The calculated t-statistic quantifies the ratio of the difference between the group means to the variability of the scores within the groups. We reported the associated p-values, indicating the probability of observing the data under the null hypothesis. A p-value below the significance level $\alpha=0.01,0.05,0.1$ leads to rejecting the null hypothesis, indicating a significant difference between the two groups.} 

\JIE{\section{Results and Analysis}\label{sec4}}

\subsection{Communication Competency of Code LLMs on HumanEvalComm (RQ1)}





\begin{figure}[h]
  \centering
  \includegraphics[width=\textwidth]{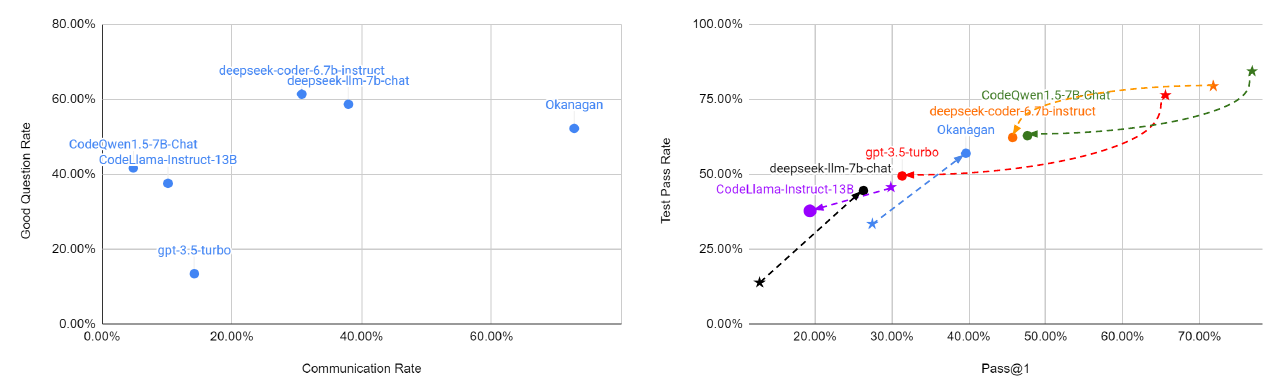}
  \caption{\JIE{Comparison of the effectiveness of the models in Communication Rate, Good Question Rate (left), and Pass@1, Test Pass Rate (right). Note that in the right figure, the stars represent the original performance of the corresponding model with the same color in the HumanEval benchmark. This shows visually how the performance has changed when the problem description is modified. }}
  \label{fig:scatter_plot}
\end{figure}

\JIE{To answer RQ1, we conducted experiments to evaluate the communication capability for problems in HumanEvalComm. \JIEW{Since we focus on the results of Code LLMs in RQ1, the results of Okanagan will be discussed separately in RQ2.} For each problem modified according to a category clarification type or combinations of clarification types, we followed the process in Figure~\ref{fig:eval_overview}. We calculated the following evaluation metrics: communication rate, good question rate, Pass@1, and Test Pass Rate. We compared Pass@1 and Test Pass Rate between the modified problem in HumanEvalComm and the original problem in HumanEval. Table~\ref{Tab:sec5_1} summarizes the overall results we generated for the evaluated models. Figure~\ref{fig:scatter_plot} rearranges the numbers in Table~\ref{Tab:sec5_1} in a visual illustration to facilitate a more direct comparison between different models. We first analyzed the overall results, then we looked into the results in each clarification category. For each category, we evaluated them with statistical testing using the Student's t-test and obtained the p-value.  }

\JIE{Let's first look at the communication rate. From Table~\ref{Tab:sec5_1} \JIEW{and Figure~\ref{fig:scatter_plot}}, the communication rate for ChatGPT, CodeLlama, and CodeQwen1.5 Chat is below 20\%, significantly less than the perfect score of 100\%. This means that for a problem description in which clarifying questions are needed for generating correct code, these models raise questions with less than a 20\% chance. The recently released DeepSeek Coder and DeepSeek Chat achieved higher communication rates of 30.76\% and 37.93\%. One hypothesis to explain this is that the general capability from DeepSeek Chat is important for a high communication rate. 
}

\JIEW{Besides communication rate, Good Question Rate is also a useful metric, because it reports the percentage of questions labeled as \textit{Good} questions based on the content of questions using an LLM-based evaluator. } In terms of Good Question Rate, likewise, \JIEW{as shown Table~\ref{Tab:sec5_1} and Figure~\ref{fig:scatter_plot}}, ChatGPT, CodeLlama, and CodeQwen1.5 Chat have a lower average question quality than DeepSeek Coder and DeepSeek Chat. Particularly, ChatGPT has a much lower rate than other open-source models. From our manual inspection, one of the reasons is because the open-source models sometimes do not follow the instructions to return either code blocks or questions. They sometimes output code blocks together with some explanations. This type of response is not a clarifying question, but the LLM-based Evaluator sometimes labels them as ``Good Question'', which we described in detail in section 3.2. Regardless, based on the numbers, there is still significant room to improve on Good Question Rate.

For pass rate measurements, ChatGPT, CodeLlama, and DeepSeek Chat achieve overall lower results than CodeQwen1.5 Chat and DeepSeek Coder for both Pass@1 and Test Pass Rate, \JIEW{based on Table~\ref{Tab:sec5_1} and Figure~\ref{fig:scatter_plot}}. The trend is similar for both Pass@1 and Test Pass Rate. \JIEW{One hypothesis is that this result is in part due to the higher Pass@1 and Test Pass Rate of CodeQwen1.5 Chat and DeepSeek Coder in the \textit{original} HumanEval benchmark. On the relative change, we see an increase in Pass@1 and Test Pass Rate from original HumanEval to HumanEvalComm for DeepSeek Chat. According to our investigation, this is because of illegal response formats: many responses from DeepSeek Chat for the original HumanEval do not have code markup, so these responses without code markup failed all the tests.} For the rest open-source models, the relative drop in the Pass@1 is between 35\% and 52\%. The relative drop in the Test Pass Rate is between 17\% and 35\%.


\begin{table*}[htbp]
\centering
\begin{tabular}{p{3cm}p{1.2cm}p{1.7cm}p{1.2cm}p{1.7cm}p{1.2cm}p{1.2cm} }
\toprule
Model & \multicolumn{2}{c}{Pass@1} & \multicolumn{2}{c}{Test Pass Rate} & Comm.  & Good  \\
  & \textit{HmEval} & \textit{HmEvalComm} & \textit{HmEval} & \textit{HmEvalComm} & Rate&Question Rate\\
 
\midrule
ChatGPT &65.58\%& 31.34\% &	76.42\%& 49.39\% & 14.21\% 	& 13.43\%           \\
CodeLlama  &29.88\%& 19.35\% &	45.71\% & 37.79\%	&10.16\%	& 37.55\%    \\
CodeQwen1.5 Chat  &76.83\%& \textbf{47.61\%} &	84.4\% & \textbf{62.89\%}	&4.82\%	& 41.68\%    \\
DeepSeek Coder &71.78\% & \textbf{45.68\%} &	79.44\% & 	\textbf{62.25\%}	&\textbf{30.76\%}	& \textbf{61.42\%}    \\
DeepSeek Chat  &12.8\%& 26.32\% &	13.86\% & 	44.52\%	&\textbf{37.93\%}	& 	\textbf{58.71\%}  \\
Okanagan {\JWR{\footnotesize (Base=ChatGPT)}} &27.45\%& \textbf{39.62\%} & 33.45\%&	\textbf{56.98\%}	& \textbf{72.73\%}	& \textbf{52.24\%}        \\
\JWR{Okanagan {\footnotesize (Base=DeepSeek Coder)}} &\JWR{21.25\%}& \JWR{\textbf{38.06\%}} &  \JWR{24.3\%}&	\JWR{\textbf{52.72\%}}	& \JWR{\textbf{82.51\%}}	& \JWR{\textbf{60.13\%}}        \\
\bottomrule
\end{tabular}
\caption{Evaluation result across all clarification categories on Pass@1, Test Pass Rate, communication rate, and Good Question Rate with different models on HumanEvalComm \JIEW{(\textit{HmEvalComm} in the table)}. Additionally, the Pass@1 and Test Pass Rate on the original problems in HumanEval \JIEW{(\textit{HmEval} in the table)} are also shown. \JIEW{Top 4 results are marked as \textbf{bold}.}}
\label{Tab:sec5_1}
\end{table*}

\begin{table*}[]
\begin{center} 
\begin{tabular}{p{1.3cm}cp{1.5cm}p{1.7cm}p{1.2cm}p{1.7cm}p{1cm}p{1.3cm}}
\toprule
Clarification  & Model & \multicolumn{2}{c}{Pass@1} & \multicolumn{2}{c}{Test Pass Rate} & Comm.  & Good   \\
 Category &  & \textit{HmEval} & \textit{HmEvalComm} \textcolor{gray}{\textit{(p-value)}} & \textit{HmEval} & \textit{HmEvalComm} \textcolor{gray}{\textit{(p-value)}} & Rate & Question Rate \\
\midrule
\multirow{7}{*}{1a} & ChatGPT & 65.58\% &  33.77\%*** & 76.42\% & 54.98\%*** & 5.84\%   &  4.55\%           \\
& & & \textcolor{gray}{\textit{(0.000)}}  &  & \textcolor{gray}{\textit{(0.000)}}    &    &         \\ \cline{2-8}
&CodeLlama & 29.88\% &  16.46\%***& 45.71\% &     36.24\%**&    13.64\%&    42.68\%
    \\
& & & \textcolor{gray}{\textit{(0.004)}}  &  & \textcolor{gray}{\textit{(0.037)}}    &    &         \\  \cline{2-8}
&CodeQwen1.5 Chat  &76.83\% &  46.34\%***& 84.4\% &    62.62\%***&    5.84\%& 43.29\%
    \\
& & & \textcolor{gray}{\textit{(0.000)}}  &  & \textcolor{gray}{\textit{(0.000)}}    &    &         \\  \cline{2-8}
&DeepSeek Coder & 71.78\% &  43.29\%***& 79.44\% &   61.2\%***& 25.97\%&    62.8\%
    \\
& & & \textcolor{gray}{\textit{(0.000)}}  &  & \textcolor{gray}{\textit{(0.000)}}    &    &         \\  \cline{2-8}
&DeepSeek Chat & 12.8\% &  21.95\%**& 13.86\% &   40.62\%***&    39.61\%&    56.71\%
    \\
& & & \textcolor{gray}{\textit{(0.029)}}  &  & \textcolor{gray}{\textit{(0.000)}}    &    &         \\  \cline{2-8}
&Okanagan & 27.45\% &  44.81\%*** & 33.45\% &     64.22\%***&  65.58\%    & 52.60\%         \\
& & & \textcolor{gray}{\textit{(0.001)}}  &  & \textcolor{gray}{\textit{(0.000)}}    &    &         \\ 
\midrule
\multirow{7}{*}{1c} & ChatGPT & 65.58\% &  53.25\%** &76.42\% &  66.37\%** & 5.84\%  &  6.49\%          \\
& & & \textcolor{gray}{\textit{(0.027)}}  &  & \textcolor{gray}{\textit{(0.028)}}    &    &         \\  \cline{2-8}
&CodeLlama & 29.88\%&  32.93\%&  45.71\% &     52.14\%&    7.79\%& 32.32\%
    \\
& & & \textcolor{gray}{\textit{(0.553)}}  &  & \textcolor{gray}{\textit{(0.172)}}    &    &         \\  \cline{2-8}
&CodeQwen1.5 Chat  &76.83\% &  67.68\%***&  84.4\% &    79.9\%***& 7.79\%& 46.95\%
    \\
& & & \textcolor{gray}{\textit{(0.000)}}  &  & \textcolor{gray}{\textit{(0.000)}}    &    &         \\  \cline{2-8}
&DeepSeek Coder & 71.78\%  &  61.59\%*&  79.44\% &   76.75\%&    15.03\%&    53.66\%
    \\
& & & \textcolor{gray}{\textit{(0.051)}}  &  & \textcolor{gray}{\textit{(0.501)}}    &    &         \\  \cline{2-8}
&DeepSeek Chat &  12.8\% &  39.63\%***&  13.86\% &   56.89\%***&    28.1\%& 61.59\%
    \\
& & & \textcolor{gray}{\textit{(0.000)}}  &  & \textcolor{gray}{\textit{(0.000)}}    &    &         \\  \cline{2-8}
&Okanagan &  27.45\% &  57.14\%*** &  33.45\% &      70.01\%*** & 55.19\%   &  42.86\%     \\
& & & \textcolor{gray}{\textit{(0.000)}}  &  & \textcolor{gray}{\textit{(0.000)}}    &    &         \\ 
\midrule
\multirow{7}{*}{1p} & ChatGPT & 65.58\% & 27.95\%*** & 76.42\%&  44.14\%*** & 31.68\%     &    26.71\%         \\
& & & \textcolor{gray}{\textit{(0.001)}}  &  & \textcolor{gray}{\textit{(0.000)}}    &    &         \\  \cline{2-8}
&CodeLlama & 29.88\% & 15.24\%***&  45.71\% &     29.41\%***&    7.74\%& 33.54\%
    \\
& & & \textcolor{gray}{\textit{(0.000)}}  &  & \textcolor{gray}{\textit{(0.000)}}    &    &         \\  \cline{2-8}
&CodeQwen1.5 Chat & 76.83\%  & 46.95\%***&  84.4\% &    59.36\%***&    1.3\%&  38.41\%
    \\
& & & \textcolor{gray}{\textit{(0.000)}}  &  & \textcolor{gray}{\textit{(0.000)}}    &    &         \\  \cline{2-8}
&DeepSeek Coder & 71.78\%   & 45.12\%***&  79.44\% &   58.57\%***&    48.7\%& 68.9\%
    \\
& & & \textcolor{gray}{\textit{(0.000)}}  &  & \textcolor{gray}{\textit{(0.000)}}    &    &         \\  \cline{2-8}
&DeepSeek Chat &  12.8\%  & 21.95\%**&  13.86\% &   43.73\%***&    37.66\%&    55.49\%
    \\
& & & \textcolor{gray}{\textit{(0.029)}}  &  & \textcolor{gray}{\textit{(0.000)}}    &    &         \\  \cline{2-8}
&Okanagan &  27.45\%  & 36.65\%* &  33.45\% &      54.16\%*** & 93.17\%   & 58.39\%       \\
& & & \textcolor{gray}{\textit{(0.082)}}  &  & \textcolor{gray}{\textit{(0.000)}}    &    &         \\ 
\midrule
\end{tabular}
\end{center}
\caption{Evaluation result for clarification categories 1a,1c,1p on Pass@1, Test Pass Rate for \JIEW{original problems in HumanEval and modified problems in HumanEvalComm}, communication rate, and Good Question Rate with different models. \JIEW{*p<0.1; **p<=0.05; ***p<0.01}}
\label{Tab:sec5_2}
\end{table*}

\textbf{Breakdown on Categories with One Clarification Type.} Besides overall results, we would like to further \JWR{understand the corresponding results within different clarification categories}. Table~\ref{Tab:sec5_2} shows the results breakdown on the clarification categories 1a, 1c, and 1p, where only one level of clarification type (\textit{Ambiguity}, \textit{Inconsistency}, and \textit{Incompleteness}) is applied to the problem. For ChatGPT, among the three clarification types, \textit{Incompleteness} has the overall highest communication rate, suppressing the communication rates of \textit{Ambiguity} and \textit{Inconsistency}. This means that \textit{Incompleteness} is relatively easier to detect and raise than \textit{Ambiguity} and \textit{Inconsistency} for models such as ChatGPT, DeepSeek Coder, and DeepSeek Chat. \textit{Inconsistency} has the lowest communication rate among the three types. One hypothesis to explain that is that \textit{Inconsistency} requires stronger reasoning capability to detect. Good Question Rate follows similar patterns as the communication rate, indicating that the quality of questions is proportional to the communication rate for Code LLMs. 

Two exceptions to the above statements are that CodeLlama and CodeQwen1.5 Chat achieved the lowest communication rate in the \textit{Incompleteness} category than in \textit{Ambiguity} and \textit{Inconsistency}.  Similar trends can be found in Good Question Rates. \JIEW{This shows that some Code LLMs such as CodeLlama and CodeQwen1.5 Chat are trained and designed in a way so that they tend to complete code rather than ask questions even when requirements are incomplete. This reflects the generative nature of LLMs: given a prompt, the LLM as a generative model essentially generates and completes text (or code in our scenario) based on the statistical model \cite{shanahan2024talking}. Thus, one hypothesis of the low result is that LLMs have disadvantages due to their generative nature when evaluating communication in coding tasks. This result also indicates that more intelligent AI agents such as LLM-based agents, where LLM as a generative model is a component, have the potential to outperform LLMs in the evaluation on communication capability \cite{shanahan2024talking}.  }

\JIEW{For the testing performance of the generated code, interestingly, \textit{Incompleteness} receives overall the lowest Pass@1 (12.8\% $\sim$ 46.95\%) and Test Pass Rate (29.41\% $\sim$ 59.36\%) for all of the models.} One hypothesis is that if no clarifying questions were asked for problems with \textit{Incompleteness}, the generated codes would be typically incorrect due to lack of information. \textit{Inconsistency} has the highest Pass@1 and Test Pass Rate, because, for problems with \textit{Inconsistency}, LLMs are sometimes able to generate correct code without asking clarifying questions. For 1a, 1c, and 1p categories, \JIEW{all except 3 changes in Pass@1 and Test Pass Rate are statistically significant, with p-values less than 0.1.}

\JIE{\textbf{Breakdown on Categories with Two Clarification Types.} Table~\ref{Tab:sec5_3} shows the results breakdown on the clarification category of 2ac, 2ap, and 2cp, where a combination of two clarification types is applied to the problem. Compared with applying one clarification type in Table~\ref{Tab:sec5_2}, two clarification types have on average slightly higher communication rates than one clarification type. This makes sense as a combination of two clarification types naturally triggers more questions than one type. Consequently, we see a similar trend for the Good Question metric.} 

\JIE{In terms of testing performance for the combination of two clarification types, both the Pass@1 and Test Pass Rate decreased significantly from one clarification type to two types. Therefore, compared with one clarification type, a combination of two clarification types further reduces the Test Pass Rate significantly, but only slightly enlarges the communication rate and the quality of clarifying questions on average. \JIEW{The slight increase in communication rate is reasonable given the increased clarification difficulty. The decreased pass rates show that it is hard for the models to get the necessary requirements for solving the task given the challenging situation for combinations of clarification types.} 2cp, with a combination of both \textit{Inconsistency} and \textit{Incompleteness}, results in lower Pass@1 and Test Pass Rate compared with 2ap and 2ac.  
For 2ac, 2cp, and 2ap categories, 75\% of the changes in Pass@1 and Test Pass Rate are statistically significant since the p-values are less than 0.05 in these changes.}

\begin{summarybox}
\textbf{Answer to RQ1:} More than 60\% of responses from Code LLMs still generate code rather than ask questions when the problem descriptions are manually modified according to different clarification categories. Typically, the Pass@1 and Test Pass Rate of Code LLMs drop by 35\% $\sim$ 52\% and by 17\% $\sim$ 35\% respectively, with statistical significance in each category for over 75\% numbers.
\\ \\
Among the three clarification types, the \textit{Incompleteness} category results in higher communication rates and Good Question Rates, but lower Pass@1 and Test Pass Rate than the \textit{Ambiguity} and \textit{Inconsistency} categories for Code LLMs. A combination of two clarification types leads to slightly higher communication rates but much lower Test Pass Rates than one clarification type. 
\end{summarybox}

\begin{table*}[]
\begin{center} 
\begin{tabular}{p{1.3cm}cp{1.5cm}p{1.7cm}p{1.2cm}p{1.7cm}p{1cm}p{1.3cm}}
\toprule
Clarification  & Model & \multicolumn{2}{c}{Pass@1} & \multicolumn{2}{c}{Test Pass Rate} & Comm.  & Good   \\
 Category &  & \textit{HmEval} & \textit{HmEvalComm} \textcolor{gray}{\textit{(p-value)}} & \textit{HmEval} & \textit{HmEvalComm} \textcolor{gray}{\textit{(p-value)}} & Rate & Question Rate \\
\midrule
\multirow{7}{*}{2ac} & ChatGPT & 65.79\% & 20.39\%*** & 76.77\% & 42.66\%*** & 5.26\%   &    7.90\%        \\ 
& & & \textcolor{gray}{\textit{(0.000)}}  &  & \textcolor{gray}{\textit{(0.000)}}    &    &         \\  \cline{2-8}
&CodeLlama  & 29.63\% & 14.2\%***  & 45.65\%&  36.95\%* &   12.5\%  &   42.59\%
    \\ 
& & & \textcolor{gray}{\textit{(0.001)}}  &  & \textcolor{gray}{\textit{(0.054)}}    &    &         \\  \cline{2-8}
&CodeQwen1.5 Chat &77.16\% & 40.12\%***  &84.28\% &  59.56\%***&   7.24\%  &   47.53\%
    \\
& & & \textcolor{gray}{\textit{(0.000)}}  &  & \textcolor{gray}{\textit{(0.000)}}    &    &         \\   \cline{2-8}
&DeepSeek Coder  &71.43\%& 40.74\%*** &79.18\%&   61.72\%*** &   26.97\% &   58.64\%
    \\
& & & \textcolor{gray}{\textit{(0.000)}}  &  & \textcolor{gray}{\textit{(0.000)}}    &    &         \\   \cline{2-8}
&DeepSeek Chat &12.96\%& 20.99\%* &14.03\%&   39.09\%*** &   44.08\% &   64.2\%
    \\
& & & \textcolor{gray}{\textit{(0.055)}}  &  & \textcolor{gray}{\textit{(0.000)}}    &    &         \\   \cline{2-8}
&Okanagan &27.15\%& 25.66\% & 33.23\%&  47.37\%*** & 64.47\%   &  45.39\%      \\
& & & \textcolor{gray}{\textit{(0.769)}}  &  & \textcolor{gray}{\textit{(0.004)}}    &    &         \\  
\midrule
\multirow{7}{*}{2cp} & ChatGPT & 77.42\% & 15.63\%***& 84.91\% & 34.79\%*** & 6.25\%   &    9.38\%        \\ 
& & & \textcolor{gray}{\textit{(0.000)}}  &  & \textcolor{gray}{\textit{(0.000)}}    &    &         \\  \cline{2-8}
&CodeLlama  & 38.24\% & 14.71\%**&60.9\%&    33.04\%*** &   9.68\%  &   29.41\%
    \\
& & & \textcolor{gray}{\textit{(0.028)}}  &  & \textcolor{gray}{\textit{(0.003)}}    &    &         \\   \cline{2-8}
&CodeQwen1.5 Chat  &73.53\%& 38.24\%*** &83.57\%&   55.82\%*** &   0\% &   29.41\%
    \\
& & & \textcolor{gray}{\textit{(0.003)}}  &  & \textcolor{gray}{\textit{(0.004)}}    &    &         \\   \cline{2-8}
&DeepSeek Coder&70.59\%& 29.41\%*** &80.93\%&   50.97\%*** &   12.9\%  &   52.94\%
    \\
& & & \textcolor{gray}{\textit{(0.000)}}  &  & \textcolor{gray}{\textit{(0.002)}}    &    &         \\   \cline{2-8}
&DeepSeek Chat &11.76\%& 26.47\% &11.76\%&   48.38\%*** &   22.58\% &   52.94\%
    \\
& & & \textcolor{gray}{\textit{(0.127)}}  &  & \textcolor{gray}{\textit{(0.000)}}    &    &         \\   \cline{2-8}
&Okanagan &32.26\%& 28.13\% & 34.98\%&  44.97\% & 84.38\%   &   59.38\%    \\
& & & \textcolor{gray}{\textit{(0.726)}}  &  & \textcolor{gray}{\textit{(0.375)}}    &    &         \\  
\midrule
\multirow{7}{*}{2ap} & ChatGPT & 59.42\% & 16.67\%*** & 71.09\% & 32.39\%***& 37.50\%  &     29.17\%     \\ 
& & & \textcolor{gray}{\textit{(0.000)}}  &  & \textcolor{gray}{\textit{(0.000)}}    &    &         \\  \cline{2-8}
&CodeLlama  &28.38\%& 17.57\%   &41.59\%&   31.27\% &   8.57\%  &   39.19\%
    \\ 
& & & \textcolor{gray}{\textit{(0.120)}}  &  & \textcolor{gray}{\textit{(0.135)}}    &    &         \\  \cline{2-8}
&CodeQwen1.5 Chat  &74.32\%& 28.38\%***  &82.71\%&   44.35\%***&   1.45\%  &   28.38\%
    \\
& & & \textcolor{gray}{\textit{(0.000)}}  &  & \textcolor{gray}{\textit{(0.000)}}    &    &         \\   \cline{2-8}
&DeepSeek Coder &71.23\%& 36.49\%***&81.36\%&   48.17\%***&   56.52\% &   70.27\%
    \\
& & & \textcolor{gray}{\textit{(0.000)}}  &  & \textcolor{gray}{\textit{(0.000)}}    &    &         \\   \cline{2-8}
&DeepSeek Chat &9.46\%& 24.32\%** &10.9\%&    35.99\%*** &   52.17\% &   56.76\%
    \\
& & & \textcolor{gray}{\textit{(0.016)}}  &  & \textcolor{gray}{\textit{(0.000)}}    &    &         \\   \cline{2-8}
&Okanagan &27.94\% & 29.17\% & 34.52\%& 43.94\% & 94.44\%   &   66.67\%    \\
& & & \textcolor{gray}{\textit{(0.874)}}  &  & \textcolor{gray}{\textit{(0.196)}}    &    &         \\  
\midrule
\end{tabular}
\end{center}
\caption{Evaluation result for clarification categories 2ac,2cp,2ap on Pass@1, Test Pass Rate for \JIEW{original problems in HumanEval and modified problems in HumanEvalComm}, communication rate, and Good Question Rate with different models. \JIEW{*p<0.1; **p<=0.05; ***p<0.01}}
\label{Tab:sec5_3}
\end{table*} 

\subsection{Comparing Okanagan with Code LLMs in communication skills (RQ2)}
\JIE{\textbf{Overview.} RQ2 aims to compare LLM agent approach, Okanagan, with the current Code LLMs in communication skills. From Table~\ref{Tab:sec5_1}, while the communication rate for ChatGPT is below 20\%, the communication rate of Okanagan with ChatGPT is over 70\%, much higher than ChatGPT and all the other models. This shows that changing from LLM to LLM agent significantly increases the communication rate. For testing performance, Okanagan achieves better results than all models except CodeQwen1.5 Chat and DeepSeek Coder in both Pass@1 and Test Pass Rate. The trend is similar for both Pass@1 and Test Pass Rate as mentioned previously. This shows the effectiveness of Okanagan in obtaining the necessary information by asking clarifying questions. }

However, one drawback is that \JWR{in the original HumanEval}, Okanagan achieves a much lower Pass@1 and Test Pass Rate than ChatGPT. This is because the multi-round structure sometimes asks questions as an initial response even for original problems, but in original HumanEval, it is expected to directly return code in the initial responses, and evaluation is conducted on the code in the initial responses. This means for original problems that do not need asking questions, Okanagan sometimes still asks questions that appear to be unnecessary, since the original problem is known as complete and has no requirement issue. This is a valid limitation and future work is required (e.g., multi-agent debate~\cite{du2023improving}) to address this limitation of asking unnecessary questions. If addressed, it indicates much stronger communication capability as LLM agent knows intelligently when to stop asking~\cite{hassan2024rethinking}. On the other hand, on HumanEvalComm, Okanagan shows that with LLM agent on top of base LLM (ChatGPT), we can get much better results in all metrics than ChatGPT. This shows the advantage in communication capability of LLM agent over LLM, as LLM agent can obtain needed info by asking questions to increase pass rates.

\textbf{Breakdown on Categories with One Clarification Type.} In terms of the results breakdown on the clarification category of 1a, 1c, and 1p, Okanagan shows a much higher communication rate than any other Code LLMs for 1a, 1c, and 1p. \textit{Incompleteness} again has the highest communication rate among the three categories, with a rate of more than 90\%. A similar trend holds for Good Question Rate, but DeepSeek Coder and DeepSeek Chat get higher Good Question Rate than Okanagan in 1a and 1c. Consequently, Okanagan achieves the top three performance for both the highest Pass@1 and Test Pass Rate among all the models with statistical significance.

\textbf{Breakdown on Categories with Two Clarification Types.} For the results breakdown on the clarification category of 2ac, 2ap, and 2cp, Okanagan has the highest communication rate among all the models, reaching a high 85\% in 2cp and 94\% in 2ap. A similar trend still holds in Good Question Rate. For Pass@1 and Test Pass Rate, Okanagan still ranks at the top 3 in most cases, but the statistical testing shows that most of the p-values are not small enough to show strong evidence of statistical significance, due to the relative smaller sample size of the categories with two types.

\JWR{\textbf{Switch Base Model in Okanagan.} We also switched base model from using ChatGPT to using another open-sourced model, DeepSeek Coder, in Okanagan to ensure a fair comparison. The reason we chose DeepSeek Coder is that it is the best model that achieves the top 4 results on all of the metrics evaluated in Table 3. The result of Okanagan (with DeepSeek Coder as base) is displayed in Table 3. }

\JWR{For HumanEvalComm dataset, compared with DeepSeek Coder, Okanagan with DeepSeek Coder as base is better in Communication Rate, similar in Good Question Rate and slightly worse in Pass@1 and Test Pass Rate. After some investigations, one potential reason of the lower pass rates is that the prompts used in Okanagan framework is suitable and effective for ChatGPT as base, but not for DeepSeek Coder as base. Because of the prompt issue, the base DeepSeek Coder sometimes doesn't follow instruction and ask question within Okanagan.} \JWR{For HumanEval dataset, Okanagan with DeepSeek Coder as base is worse than DeepSeek Coder. This is similar to Okanagan (with ChatGPT as base) comparing with ChatGPT, due to the limitation of Okanagan that sometimes asks unnecessary questions, which has been discussed in the paper.}

\JWR{Comparing Okanagan with ChatGPT as base and DeepSeek Coder as base, Okanagan with DeepSeek Coder as base is worse in Test Pass rates and Pass@1 for HumanEval and HumanEvalComm, but better in Communication Rate and Good Question Rate. The former could be because DeepSeek Code as base is less effective than ChatGPT as base in prompt understanding within Okanagan. The latter could be due to the fact that DeepSeek Coder is better than ChatGPT in Communication Rate and Good Question Rate. }

\begin{summarybox}
\textbf{Answer to RQ2:} Okanagan, as a LLM agent approach that uses LLM (specifically ChatGPT), effectively increases Communication Rate and Good Question Rate by an absolute 58\% and 38\% respectively, and thus boosts Pass@1 and Test Pass Rate by an absolute 8\% and 7\% respectively, when the problem descriptions are modified based on given clarification categories. \JWR{However, Okanagan tends to still ask questions that appear to be unnecessary for original problems that do not need questions. This indicates headroom for achieving more effective communication capability using LLM agent, and the need to strike a balance between asking unnecessary questions and truly needed questions.} 
\end{summarybox}

\subsection{\JWR{Manual Evaluation of LLM-based Evaluator (RQ3)}}

\JWR{\textbf{Overview.} We conducted a manual evaluation of LLM-based evaluator to answer RQ3.  We recruited six graduate students at the University of British Columbia to conduct the manual evaluation. The 6 students are from the software engineering lab in the computer science department, all having at least two years of professional software development background. We asked the students to manually assess the results of LLM-based evaluator and manually mark the quality of models' responses within one week.}

\JWR{\textbf{Manual Evaluation Setup.} We selected a subset of problem samples in HumanEvalComm using \textit{representative} sample size. We have around 4300 samples in total (for 6 models) available for manual evaluation. However, as manual evaluation is time-consuming for the recruited students, we used sample size calculator to select 60 samples for each model, which is statistically representing our population size. In the sample size calculator, we set Confidence Level as 95\%, Margin of Error as 5\%, and
Population Proportion as 50\%. We used a hash function to sample 60 problems from HumanEvalComm for each of the 6 models, and created a guideline, with examples detailing the instructions for the students to do the manual evaluation. To avoid introducing bias, we hide the actual names of the 6 models and used model 1 to model 6 instead of actual model names. For each sample of a give model, we asked two students to do the labels independently. The goal is to conduct human evaluation to evaluate the quality of clarifying questions from models (i.e., is question asked or not, what's the question quality) and answers from our LLM evaluator to answer the questions (i.e., what's the answer quality). Concretely, the tasks we asked the students to rate for each sample is the following: }
\begin{itemize}
    \item \JWR{\textbf{Question or Not}: The students were assigned to assess is question asked or not by the model given the problem description. The labels are 1: Yes, 0: No.}
    \item \JWR{\textbf{Question Quality}: We then asked students to label the quality of questions. Similar to the ones in Good Question Rate, the labels are 3: Good (The model asks clarifying questions that help recover the missing info), 2: Fair (The model asks OK, reasonable clarifying questions, but the questions may not cover the missing info), 1: Bad (The model asks no questions or completely irrelevant questions that do not help at all to recover the missing/clarifying information). }
    \item \JWR{\textbf{Answer Quality}: We asked students to also label the quality of answers to the questions. The labels are 3: Good (The answer recovers the missing info), 2: Fair (The answer is OK and reasonable, but may not recover the missing info), 1: Bad (The answer does not make sense at all), -1: False Recovery (The answer recovers the missing info, although `response' doesn't have questions), 0: No Recovery (The answer does not recover the missing info if `response' doesn't have questions. Note that 1,2 or 3 is selected if the model `response' does have questions, otherwise, -1 or 0 is selected.) }
\end{itemize}

\begin{table}
    \begin{tabular}{p{2.5cm}p{3cm}p{3cm}p{3cm}p{1cm}}
        \toprule
         \textbf{model} & \textbf{Good Answer Rate (questions)} & \textbf{Acceptable Answer Rate (questions)} & \textbf{False Recovery Rate (non-questions)} & \textbf{Kappa \(\kappa\)}  \\
        \midrule
        ChatGPT  & 50.0\% & 50.0\% & 0.94\%& 0.85  \\
        CodeLlama & 100\% & 100\% & 32.5\%& 0.96  \\
        CodeQwen1.5 Chat & 100\% & 100\% & 40.7\% & 0.72  \\
        DeepSeek Coder & 69.2\% & 100\% & 38.0\% & 0.92  \\
        DeepSeek Chat & 41.91\% & 59.0\% & 34.8\%& 0.77  \\
        Okanagan & 31.5\%& 53.3\%& 0\% & 0.77  \\
        \bottomrule
    \end{tabular}
    \caption{\JWR{Metrics from Manual Evaluation for Answers from LLM-based Evaluator To Models' Response. Each row shows different metrics for LLevaluator's answers to the model's responses, with the Kappa (\(\kappa\)) value.}}
    \label{tab:manaul_eval_answer_results}
\end{table}

\JWR{We then converted the above labels to 4 metrics and reported them: }

\begin{itemize}
    \item \JWR{\textbf{Communication Rate}: The percentage of model responses that ask clarifying questions.}
    \item \JWR{\textbf{Good Question Rate}: The percentage of model responses asking questions with Good question quality label (3).}
    \item \JWR{\textbf{Good Answer Rate (questions)}: The percentage of model's Good answer to questions, where the answer has a Good (3) answer quality label. This is for model responses that are questions.}
    \item \JWR{\textbf{Acceptable Answer Rate (questions)}: The percentage of model asking questions with Good (3) and Fair (2) question quality label. This is for model responses that are questions.}
    \item \JWR{\textbf{False Recovery Rate (non-questions)}: For model responses that do not have questions, False Recovery Rate is the percentage of the answers that recovers the missing info among all the answers to model responses that do not have questions. Ideally, if the model responds with non-questions, we don't want to provide answer that recovers the missing info, and thus falsely increases the test pass rates and Pass@1 in our evaluation.}
\end{itemize}

\JWR{\textbf{Manual Evaluation Results.} Table~\ref{tab:manaul_eval_answer_results} shows the comparison of the performance of the LLM-based evaluator's answers to various models across three metrics: Good Answer Rate, Acceptable Answer Rate, and False Recovery Rate, along with their inter-rater reliability represented by Cohen's Kappa \(\kappa\). the results of each metric from the recruited students. Note that each result has two student raters, so the average and the Kappa are reported in the three rates columns and Kappa column respectively. The Kappa is between 0.7-1.0 for all the rows in the result, indicating that two raters reach substantial or near-perfect agreements.  Overall, all the models show at least over 50\% Acceptable Answer Rate. All models except DeepSeek Chat and Okanagan also show over 50\% Good Answer Rate. This illustrates the LLM-based evaluator's ability to provide correct or acceptable answers to the questions posed by different models. In terms of False Recovery Rate, ChatGPT and Okanagan show low False Recovery Rates, with 0.94\% and 0\%, respectively, but the other models show higher False Recovery Rates, between 30\% and 40\%.  This indicates the potential risk for those models that the increased test pass rates may not be contributed from the model's ability to ask questions. Future work should be spent on further reducing the False Recovery Rates as much as possible. }

\begin{figure}[h]
  \centering
  \includegraphics[width=\textwidth]{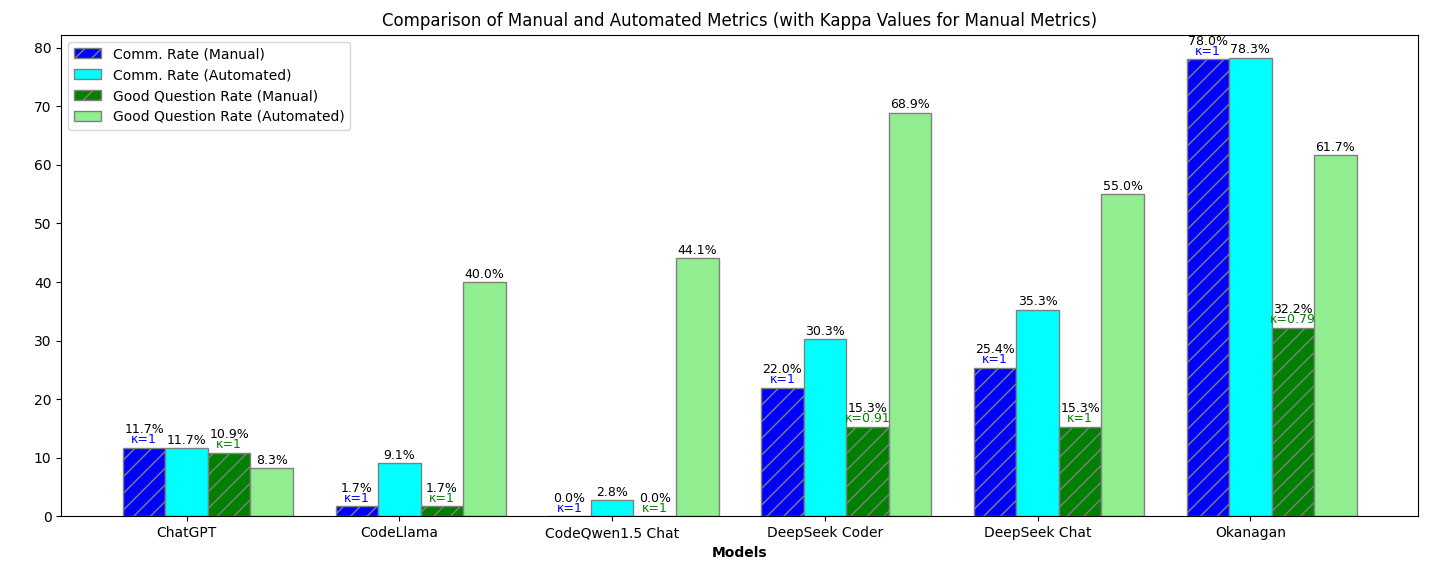}
  \caption{\JWR{Comparison of Comm. Rate and Good Question Rate between Manual Evaluation and Automated Evaluation Across Models. Each row shows the resulting percentage of a model on a particular metric followed by the Kappa (\(\kappa\)) value in parentheses.}}
  \label{fig:manual_eval}
\end{figure}

\JWR{\textbf{Comparing metrics between Automated and Manual Evaluation.} To measure the reliance of the metrics used in our evaluation, we further compared the difference between the metrics in our automated evaluation and the same metrics obtained from the manual evaluation. We calculated the automated metrics, Communication Rates and Good Question Rate, on the same data we used for the manual evaluation. Figure~\ref{fig:manual_eval} shows the direct comparison of Communication Rate and Good Question Rate between manual and automated evaluation, respectively. For manual metrics, as aforementioned, the average of two student raters and the Kappa are reported in Figure~\ref{fig:manual_eval}. }

\JWR{For communication rate, the automated metrics align well in general with the manual metrics across different models, with a difference of less than 10\%. The alignment is excellent for ChatGPT, Okanagan, CodQwen1.5 Chat with less than 3\% difference. On the other hand, there appear to be discrepancies in Good Question Rate, especially for models such as CodeLlama, CodeQwen1.5 Chat, and DeepSeek Coder.  This could be due to differences in the criteria used by humans and automated systems to judge questions as ``good''. The recruited students have a clearer and higher bar to determine a ``good'' question, while the automated LLM-based evaluator tends to label more good questions that should have been labelled fair/bad questions.  }

\begin{summarybox}
\JWR{\textbf{Answer to RQ3:} The LLM-based evaluator provides acceptable answers to models' responses, with higher than 50\% Acceptable Answer Rate for all models and Good Answer Rate  for most models. False Recovery Rates need to be further reduced for some models to ensure the reliability of test pass rates and Pass@1. The Communication Rate in automated evaluation aligns reasonably well with the manual evaluation, but for the Good Question Rate, LLM-based evaluator tends to mark more ``Good'' questions than it should. Overall, the two metrics can provide useful insights to guide our experiment, but they may not always align with human judgment, particularly for subjective criteria like Good Question Rate.}
\end{summarybox}

\subsection{\JWR{Investigating Different Impacts of Prompt Strategies and Hyperparameters (RQ4)}}

\subsubsection{Investigating Different Extent of Incomplete Modification (1p)}\label{sec:diff_1p} 
To dig deeper into the results of incomplete modification (1p), we did further investigations to \JWR{understand the corresponding results for different ratio of removed content in the problem description of 1p.} We investigated the results by removing a random text block in the problem descriptions. Specifically, for each problem description in words, we randomly remove a list of consecutive words where the size of the list is $X$ percentage of the total number of words in the problem description. We empirically choose the values $X=30\%, 50\%, 90\%$ in this experiment. Table~\ref{Tab:dataset_example} shows an example of problem descriptions with different $X$. We can see that even when $X$ is as small as 30\%, it would be ideal for LLM to ask questions to get the missing piece of information to ensure the LLM fully understands the problem and generates high-quality code. When $X$ becomes larger, such as 90\%, the problem description becomes almost impossible to conduct the code generation task with high accuracy. 

To mitigate the risk of randomness in this investigation, for each problem, we ran the experiment 5 times, and reported the metrics used in~\cite{ouyang2023llm}, including Mean, Variance, Max Diff, and Ratio of Worst. We calculated the mean and variance of the 5 Test Pass Rates and communication rates for each problem, and reported the average among all problems, as Mean and Variance. The Max Diff is the maximum value of the maximum diff among all problems.  ``Ratio of Worst (Cases)'' is the ratio of problems with the maximum diff of test pass rate being 1. Please refer to~\cite{ouyang2023llm} for complete descriptions of these metrics. We report the Test Pass Rates and communication rates in Table~\ref{Tab:rq1} based on the percentage of removed descriptions (0\%, 30\%, 50\%, and 90\%). We used ChatGPT 3.5 as the model in the investigation. We can see that, in terms of Test Pass Rates, as the percentage of removed information increases, there is a noticeable decrease in the mean Test Pass Rate, indicating that incomplete problem descriptions negatively affect the ability to pass tests. This trend is further supported by the variance and maximum difference metrics, which show increasing variability and differences in Test Pass Rates as information is removed. The ratio of the worst case also suggests that a higher percentage of removed information leads to a lower Test Pass Rate. 

\begin{figure}[h]
  \centering
  \includegraphics[width=.5\textwidth]{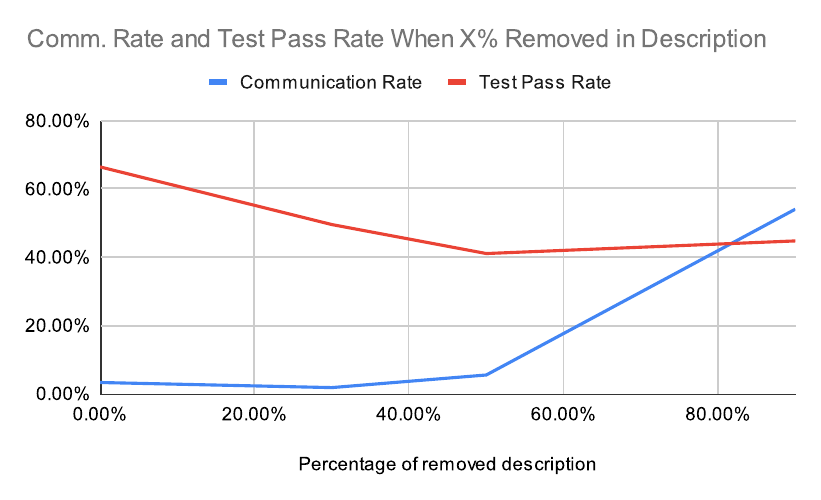}
  \caption{\JIE{The chart of communication rate and Test Pass Rate when different X\% of content is removed in the problem description.}}
  \label{fig:remove_chart}
\end{figure}

For communication rates, incomplete problem descriptions lead to an increase in the mean communication rate as the percentage of removed information increases. This is expected because when there is more missing information, LLM tends to ask more questions rather than directly generating code. The variance and maximum difference metrics also reflect higher variability and differences in communication rates with incomplete problem descriptions. The ratio of the worst case indicates that a higher percentage of removed information results in a more significant increase in communication rates. Figure~\ref{fig:remove_chart} shows a visual comparison of these two metrics using numbers from Table~\ref{Tab:rq1} when different X\% of content is removed in the problem description. This shows visually that LLM tends to ask more questions as more content in the description is removed, but this starts to happen only \text{after} half of the descriptions are removed. When 90\% of the description is removed, LLM asks questions for only 54\% of problem descriptions. To summarize, 95\% of responses from Code LLMs still generate code even when 50\% of problem descriptions are randomly removed. When the removed percentage of description increases to 90\%, 46\% of responses from Code LLMs still generate code. This shows a rather weak ability of Code LLMs to ask clarifying questions when information is randomly removed and therefore indicates plenty of research opportunities in pushing the curves of LLM or LLM agent toward the human software engineers.

\subsubsection{Investigating Different Hyperparameters}
\JIE{To evaluate the impact of the hyperparameter in the experiments, we also investigated using different temperatures as the hyperparameter of ChatGPT. Temperature is a hyperparameter that controls the randomness of ChatGPT-generated text. The default temperature of ChatGPT is 1.0 and we tested the result of using temperature as 0, 1.0, 2.0. Same as in the previous incomplete modification investigation, we ran the experiment 5 times and checked the metrics including mean and variance. We ran with the percentage of removed information being 50\%. }

\JIEW{Table~\ref{Tab:rq3} presents results on the impact of the temperature hyperparameter in ChatGPT in HumanEvalComm.} We found that the mean Test Pass Rate dropped from 49.6\% (variance=0.088, max diff=0.523, ratio of worst=0.390) to 40.7\% (variance=max diff=ratio of worst=0), and the communication rate changed from 1.8\% (variance=0.011,max diff=0.067, ratio of worst=0.067) to 3.7\% (variance=max diff=ratio of worst=0) when the temperature changed from 1 to 0. We also tested 2.0 as temperature, but found that most of these requests timed out due to longer processing time on the OpenAI server end due to high temperature. We can see that the variance, max diff, and ratio of worst become 0 when the temperature is 0. This means lower temperature does indicate much more deterministic and focused results. Interestingly, as the temperature dropped from 1 to 0, the mean Test Pass Rate dropped, but the communication rate increased. This could be because lower temperature leads to less creative and diverse output to ``guess'' the code, therefore somehow forcing the model to ask questions to seek additional information. \JIEW{Given the temperature of 1.0 is the default setting, and the results do not show a significant impact of varying temperature, in the evaluation, we use the temperature of 1.0 for both ChatGPT and Okanagan.}

\subsubsection{\JWR{Investigating Different Prompts}}
\JWR{To evaluate the impact of prompt engineering, we evaluated the results using different prompt templates and strategies. Table ~\ref{tab:apx_prompts_comm} and Table~\ref{tab:apx_prompts_ok} in Appendix shows different prompt templates in HumanEvalComm evaluation and Okanagan, respectively. We designed the experiments to make prompt variants for two major prompts in the evaluation: 1) the initial prompt (first row in Table~\ref{tab:apx_prompts_comm}) to prompt the model to either ask question or generate code, and 2) the prompt for LLM-based evaluator (second row in Table~\ref{tab:apx_prompts_comm}) to provide answer to model's questions. The prompt variants are displayed in Table~\ref{Tab:apx_prompt_variants}.}

\JWR{Table~\ref{tab:prompt_eng} shows the results of different prompt variants in the evaluation. We chose ChatGPT as a representative model from LLM, and Okanagan from LLM agent. Based on the results, we can see that when using 1st prompt variant and 1st prompt one-shot, the Good Question Rate is affected, with 22\% absolute increase for ChatGPT and -3\% decrease for Okanagan. This shows that Good Question Rate can be easily affected by prompt engineering, which aligns with our finding on the unreliability of this metric in RQ3. On the other hand, the impact of these prompt variants are less severe, typically within 10\%, in Communication Rate (-2\% to +13\%), Pass@1 (0\% to +9\%) and Test Pass Rate (+2\% to +11\%). Hence, future work should focus on how to reduce the impact of prompt engineering for this type of evaluation. We did not run the experiments for the prompt variant and one-shot prompt variant for the LLM-based evaluator prompt due to cost limitation based on our experiments in the first prompt variants. }

\begin{table*}[]
\begin{center} 
\begin{tabular}{cccccc}
\toprule
\% Removed & Metric   &  Mean & Variance & Max Diff & Ratio of Worst \\
\midrule
0\% & Test Pass Rate        & 66.5\%    &0.122  &0.690  &0.561      \\
\midrule
0\% & Communication Rate        & 3.3\% &0.019  &0.110  &0.110       \\
\midrule
30\% & Test Pass Rate       & 49.6\%     &     0.088    &   0.532 & 0.390        \\
\midrule
30\% & Communication Rate       & 1.8\% &0.011  &0.067  &0.067       \\
\midrule
50\% & Test Pass Rate       & 41.1\%    &0.085  &0.523  &0.317        \\
\midrule
50\% & Communication Rate       & 5.5\% &0.028  &0.165  &0.165       \\
\midrule
90\% & Test Pass Rate       & 44.8\%    &0.124  &0.684  &0.543       \\
\midrule
90\% & Communication Rate       & 54.1\%    &0.120  &0.604  &0.604     \\
\bottomrule
\end{tabular}
\end{center}
\caption{Results on the average values of Test Pass Rates and communication rates with different percentages of content removed in the problem descriptions. \JIE{Due to the randomness involved, the experiment was run 5 times and metrics used in~\cite{ouyang2023llm} were reported.} }
\label{Tab:rq1}
\end{table*}

\begin{table*}[]
\begin{center} 
\begin{tabular}{ccccccc}
\toprule
\% Removed & Temperature & Category	&  Mean & Variance & Max Diff & Ratio of Worst \\
\midrule
50\% & 1& Test Pass Rate		& 0.496     &     0.088    &   0.532 & 0.390        \\
\midrule
50\% & 1& Communication Rate		& 0.018	&0.011	&0.067	&0.067       \\
\midrule
50\% & 0& Test Pass Rate		& 0.407	&0 	&0	&0        \\
\midrule
50\% & 0& Communication Rate		& 0.037	&0	&0	&0       \\
\bottomrule
\end{tabular}
\end{center}
\caption{Results on the average values of test pass rates and communication rates with different temperatures as the hyperparameter of LLM in  HumanEvalComm.}
\label{Tab:rq3}
\end{table*}

\begin{table}[ht]
    \centering
    \begin{tabular}{p{1.5cm}p{1.5cm}p{2.2cm}p{2.2cm}p{2.2cm}p{2.2cm}}
        \toprule
         \textbf{Prompt} & \textbf{Model} & \textbf{Comm. Rate} & \textbf{Good Question Rate} & \textbf{Pass@1} & \textbf{Test Pass Rate} \\
        \midrule
        Original Prompt & ChatGPT &  14.21\% & 13.43\% & 31.34\% & 49.39\% \\
        1st Prompt Variant & ChatGPT &  27.24\% {\textcolor{teal}{(+13.03\%)}} & 36.19\% {\textcolor{teal}{(+22.76\%)}} & 41.12\% {\textcolor{teal}{(+9.78\%)}} & 60.45\% {\textcolor{teal}{(+11.06\%)}} \\
        1st Prompt One-Shot & ChatGPT &  17.68\% {\textcolor{teal}{(+3.47\%)}} & 36.13\% {\textcolor{teal}{(+22.70\%)}} & 38.84\% {\textcolor{teal}{(+7.50\%)}} & 58.61\% {\textcolor{teal}{(+9.22\%)}} \\
        Original Prompt & Okanagan & 72.73\% & 52.24\% & 39.62\% & 56.98\% \\
        1st Prompt One-Shot & Okanagan & 70.56\% {\textcolor{red}{(-2.17\%)}} & 48.77\% {\textcolor{red}{(-3.47\%)}} & 40.34\% {\textcolor{teal}{(+0.72\%)}} & 59.38\% {\textcolor{teal}{(+2.40\%)}} \\
        \bottomrule
    \end{tabular}
    \caption{\JWR{Results of Different Prompt Variants for the 1st Prompt in Evaluation for Selected Models. Each row shows the resulting metrics of a model for evaluation metrics.}}
    \label{tab:prompt_eng}
\end{table}

\begin{table*}
\begin{center} 
\begin{tabular}{p{2cm}p{8cm}}
\toprule
\% Removed in Description   & Example Problem Description \\
\midrule
0\% &    def encode\textunderscore cyclic(s: str):  returns encoded string by cycling groups of three characters.    \# split string to groups. Each of length 3.    groups = [s[(3 * i):min((3 * i + 3), len(s))] for i in range((len(s) + 2) // 3)]    \# cycle elements in each group. Unless group has fewer elements than 3.    groups = [(group[1:] + group[0]) if len(group) == 3 else group for group in groups]    return "".join(groups) def decode\textunderscore cyclic(s: str):    takes as input string encoded with encode\textunderscore cyclic function. Returns decoded string.    \\
\midrule
30\% &   def encode\textunderscore cyclic(s: str): returns encoded string by cycling groups of three characters. split string to in each group. Unless group has fewer elements than 3. groups = [(group[1:] + group[0]) if len(group) == 3 else group for group in groups] return "".join(groups) def decode\textunderscore cyclic(s: str): takes as input string encoded with encode\textunderscore cyclic function. Returns decoded string.      \\
\midrule
50\% &  def encode \textunderscore cyclic(s: str): returns encoded string by cycling groups of three characters.  split string to groups. Each of length 3. groups = [s[(3 * i):min((3 * i + takes as input string encoded with encode\textunderscore cyclic function. Returns decoded string.     \\
\midrule
90\% &   def encode\textunderscore cyclic(s: str): encode\textunderscore cyclic function. Returns decoded string.   \\
\bottomrule
\end{tabular}
\end{center}
\caption{Example of \FHF{randomly removing parts of the problem description from problem number 38 of HumanEval}.}
\label{Tab:dataset_example}
\end{table*}


\begin{table*}
\centering
\begin{center} 
\begin{tabular}{p{3cm}p{10cm}}
\toprule
1st Model Input &
You are an expert software developer who writes high-quality code. With below information, please either generate Python3 code (Respond directly with code only with markdown), or ask clarifying questions: 
\begin{python}
def how_many_times(string: str, substring: str) -> int:
    """ 
    Find how many times a given substring can be found
    in the original string.
    """
\end{python} \\ \midrule
1st Model Response
&
Sure, I can help you with that. However, I need more information to provide a precise solution. Could you please provide the following details?
1. What should be the behavior if the substring is not found in the string?
2. What should be the behavior if the substring is an empty string?
3. Are there any specific constraints or requirements you would like to consider?
With these details, I can provide a more accurate solution.
 \\ \midrule
LLM-based Evaluator Output &
Answer to Questions: 1. If the substring is not found in the string, the function should return 0.
2. If the substring is an empty string, the function should return 0.
3. There are no specific constraints or requirements other than counting the overlapping cases.
 \\ \midrule
2nd Model Input &[1st Model Input][1st Model Response][Answer to Questions]... Given the above conversations, generate Python code directly (Markdown) to solve the coding problem:
 \\ \midrule
 2nd Model Response
 &
 Sure, based on the provided information, here is a Python function that should solve the problem:
 \begin{python}
 def how_many_times(string: str, substring: str) -> int:
    """
    Count the number of non-overlapping occurrences of a 
    substring in a string.

    Parameters:
    string (str): The string to search within.
    substring (str): The substring to search for.

    Returns:
    int: The number of times the substring occurs in the 
    string, or 0 if the substring is empty.
    """
    if not substring:
        return 0

    count = 0
    start = 0

    while start < len(string):
        pos = string.find(substring, start)
        if pos != -1:
            count += 1
             # Ensure non-overlapping occurrences
            start = pos + len(substring) 
        else:
            break

    return count

 \end{python}
 
 \\ \bottomrule
\end{tabular}
\end{center}
\caption{\JIEW{Example case for DeepSeek Coder in problem 18 of HumanEvalComm. The clarification type of this problem is 1p (\textit{Incompleteness}). The communication rate is 100\%, the Good Question Rate is \textbf{Good}, and Test Pass Rate is 100\% for this example. }}
\label{Tab:sec5_demo1}
\end{table*}

\begin{table*}
\centering
\begin{center} 
\begin{tabular}{p{4cm}p{11cm}}
\toprule
1st Model Input &
You are an expert software developer who writes high quality code. With below information, please either generate Python3 code (Respond directly with code only with markdown), or ask clarifying questions: 
\begin{python}
def triangle_area(a, h):
    """Given length of a side and high return area for a triangle.
    >>> triangle_area(5, 3)
    15
    """
\end{python} \\ \midrule
1st Model Response
&
  Here is the Python code for the function ``triangle\_area'':
\begin{python}
def triangle_area(a, h):
    return (a * h) / 2
\end{python}
This function takes two arguments, ``a'' and ``h'', which represent the length of a side and the height of a triangle, respectively. It returns the area of the triangle, which is calculated by multiplying the length of the side by the height and then dividing by 2.

Here is an example of how to use the function:
\begin{python}
print(triangle_area(5, 3))  # Output: 15
\end{python}
This will print the area of a triangle with a side length of 5 and a height of 3, which is 15.
 \\ \midrule
LLM-based Evaluator Output &
Answer to Questions: N/A  
 \\ \midrule
2nd Model Input &[1st Model Input][1st Model Response][Answer to Questions]... 
      Given the above conversations, generate Python code directly (Markdown) to solve the coding problem:
 \\ \midrule
 2nd Model Response
 &
 Given the conversations, here is the Python code to solve the coding problem:
 \begin{python}
 def triangle_area(a, h):
 return (a * h) / 2
 
 print(triangle_area(5, 3))  # Output: 15
 \end{python}
 This code defines a function ``triangle\_area'' that takes two arguments, ``a'' and ``h'', which represent the length of a side and the height of a triangle, respectively. The function returns the area of the triangle, which is calculated by multiplying the length of the side by the height and then dividing by 2.

 The code also includes a call to the ``triangle\_area'' function with the arguments ``5'' and ``3'', which prints the area of the triangle, which is `15'.

 \\ \bottomrule
\end{tabular}
\end{center}
\caption{\JIE{Example case for CodeLlama in problem 45 of HumanEvalComm. The clarification type of this problem is 1c (\textit{Inconsistency}). The communication rate is 100\% given both code block and text are returned rather than code block only. The Good Question Rate is \textbf{Bad}, indicating no clarifying questions are detected.} }
\label{Tab:sec5_demo2}
\end{table*}

\begin{summarybox}
\JWR{\textbf{Answer to RQ4:} while LLMs begin to ask more clarifying questions after 50\% of the description is removed, even with 90\% removed, only 54\% of responses from LLMs asked questions. Variations in temperature hyper-parameter did not significantly affect performance. On the other hand, prompt engineering has a larger impact on Good Question Rate, but less impact (within 10\% absolute change) on Communication Rate, Pass@1 and Test Pass Rate. }
\end{summarybox}

\JWR{\section{Discussion}}
\label{sec6}
\JWR{In this section, we provide some example cases to understand the results of Code LLMs and Okanagan. Then we also add discussions on 1) why not using automated modification, and 2) how does LLM-based evaluator impact the evaluation results. Finally, we share the implications and future work of this study.}  
\subsection{Example Cases}
\label{sec:examples}
Besides the numeric results, we also study two cases to better understand the evaluation steps. Table~\ref{Tab:sec5_demo1} illustrates the result for a successful example problem (HumanEval/18) in HumanEvalComm. The clarification type of this problem is 1p (\textit{Incompleteness}), so the requirement of \textit{``Count overlapping cases.''} is missing in the problem description. Given this 1st model input, DeepSeek Coder was able to detect the incomplete requirement in the problem and ask clarifying questions rather than generating code directly. Therefore, the communication rate is 100\% for this single case. With the 1st model input and response, the LLM-based evaluator outputs the Good Question Rate of \textbf{Good} for our evaluation and answers to the clarifying questions. This answer provides the missing requirement of ``counting overlapping cases'' from the original problem. 

In the 2nd model input, the model is given the previous conversations, including the 1st model input, 1st model response (clarifying questions), and the answer from the LLM-based evaluator to the clarifying questions. With the requirement of ``counting overlapping cases'', the model correctly solved the problem with 100\% Test Pass Rate. Note that we can again let the model either ask questions or generate code, same as the instruction in the 1st model input, but we chose to perform only one question-answer round for simplicity of our evaluation.

\JIE{Another case is shown in Table~\ref{Tab:sec5_demo2} for another example problem (HumanEval/45) in HumanEvalComm using CodeLlama. The clarification type of this problem is 1c (\textit{Inconsistency}): in the test case, triangle\_area (5, 3) returns 15 instead of 7.5. This causes \textit{Inconsistency} between the test case and the problem description. Given the 1st model input, CodeLlama did not return a code block only according to the instruction, but a mix of code block and text explanation. Therefore, the communication rate is 100\% for this case, but the LLM-based evaluator outputs the Good Question Rate of 1 (No questions). In this case, the \textit{Inconsistency} issue was not captured and no clarifying questions were asked. For the evaluation metrics, the Good Question Rate of \textbf{Bad} successfully punished this case, but Test Pass Rate and communication rate failed to capture and punish the issue.}

\JWR{\subsection{Why not Automated Modification?}\label{sec:auto_modification}}

\JWR{A natural question is that whether it's possible to use automated modification instead of manual modification which requires efforts to hand-craft a dataset. Before we started manual modification, it should be noted that we tried to use LLM to modify the problem description, but we found that the modification by LLMs did not meet our standard. Specifically, the modification from using LLMs cannot guarantee that the modification will trigger clarifying questions. Similar limitations on using LLMs for Requirement Engineering have been also reported in~\cite{arvidsson2023prompt}.}

\JWR{We designed a prompt that generated auto-modified problems for all of 164 problems in HumanEval. 
We have uploaded the results for the 164 problems in HumanEval at \href{https://github.com/jie-jw-wu/human-eval-comm/tree/main/auto_modified_problems}{\uline{this link}}.
Table~\ref{tab:auto_modified_problems} show the result of two examples. We can see that in the first example,  ``a positive floating point number'' is
changed to ``a number'', with the intention to make
the problem ambiguous. Although this modification
is made, it does not trigger clarifying questions be-
fore generating correct code, which is required for
crafting modified problems in this study. In the second example, a new test case is added to make
the problem inconsistent, but this modification does
not create any conflict or inconsistency.
Hence, we chose to manually modify all of the problem descriptions to provide this guarantee for HumanEvalComm. }

\JWR{\subsection{How does LLM-based Evaluator Impact Results of Okanagan and Other Models?}}
\JWR{We further discuss about some models that get higher rates than Okanagan in metrics such as pass rates and Good Question Rates in Table~\ref{Tab:sec5_1}, from the LLM-based evaluator perspective. For Good Question Rate, DeepSeek Coder and DeepSeek Chat get comparable or higher rates than Okanagan with ChatGPT (or DeepSeek Code) as base. One potential cause of this is because the LLM-based evaluator tends to  mark more ``Good'' questions than it should for DeepSeek Coder and DeepSeek Chat, as illustrated in Figure~\ref{fig:manual_eval}. For Test Pass Rate and Pass@1, CodeQwen 1.5 Chat and DeepSeek Coder achieve better results than Okanagan with ChatGPT (or DeepSeek Code) as base. One of the contributors is that False Recovery Rate for CodeQwen 1.5 Chat and DeepSeek Coder is much higher than Okanagan. This could cause the issue that even though CodeQwen 1.5 Chat and DeepSeek Coder do not ask any relevant questions, the evaluator still sends them relevant information that results in higher Test Pass Rate and Pass@1. Future work is needed to reduce the bias introduced by LLM-based evaluator. }

\JWR{\subsection{How does other LLM-based Agent work?}}
\JWR{ To enable a more complete comparison, we implemented and evaluated AgentCoder~\cite{huang2023agentcoder}. AgentCoder~\cite{huang2023agentcoder} is one of the leading agent-based code generation pipelines, with reported pass@1 scores on the HumanEval benchmark of 96.3 using GPT4 and 79.9 using chatGPT, making it ideal for use as a baseline for this reasearch. }

\JWR{In implementation, we made some code changes in AgentCoder to make it work for our scenario in HumanEvalComm. In round 1 of our evaluation, we ran AgentCoder, asking the model to either generate code or clarifying questions. For this step, we had to change the functioning of AgentCoder. Originally made to run on the HumanEval dataset, we run AgentCoder on the new HumanEvalComm dataset, which has modified prompts asking the model to generate clarifying questions. Due to this, for round 1, we use only \textit{programmer} component from AgentCoder since the response is not limited to code but to clarifying questions. This leads to the generation of a completion list that contains either clarifying questions or code. For round 2, we generate answers to these questions. For round 3, we run the entirety of AgentCoder, with all the context from the previous two rounds. We published our implementation to our github repo in \href{https://github.com/jie-jw-wu/human-eval-comm/blob/main/README_AgentFramework.md}{\uline{this link}} that includes the details about our implementation and changes made. }
\JWR{However, we found that none (0\%) of AgentCoder's responses asks clarifying questions, although we tried our best to adapt AgentCoder for HumanEvalComm that generates code with clarifying questions. This shows the LLM agents usually have specialized and customized tasks, thus making it difficult to be directly adapted for our task of asking clarifying questions in HumanEvalComm. Future work is needed to more easily adapt LLM agent methods in similar benchmarks.  }

\JWRR{\subsection{Limitations}}
\JWRR{We have done our best to conduct evaluations of different models in our work. However, we have also identified a few limitations throughout this research. The limitations are summarized as follows.}

\JWRR{\textbf{High demand for human resources in modifications.} As mentioned in Section \ref{sec:auto_modification}, using automated techniques has a much lower quality than human-curated approach. Therefore, to ensure the highest possible standard for HumanEvalComm dataset, we chose to manually curate the dataset. However, this manual process of modifying problem descriptions across various categories requires significant labor costs and a heavy reliance on human expertise, which can be particularly challenging for research environments with limited resources.  Relying on experienced software engineers to review and modify each problem significantly increases the costs and duration of research. It is worth mentioning that other researchers do not require manual effort to use HumanEvalComm. So, the resource-constraint research groups can just use our benchmark as-is for testing their models.}

\JWRR{\textbf{Subjectivity in manually modified problems.}  While manual modification was necessary to ensure the modifications would trigger clarifying questions, different engineers might approach the same problem differently. Although we have used specific guidelines to modify the problem sets as mentioned in Section \ref{sec2}, given the vast potential space for modifications, the subjectivity of engineers may still affect the quality and standardization of the modified problems in HumanEvalComm.}


\JWRR{\textbf{Okanagan tends to ask unnecessary questions.} While Okanagan demonstrated improved communication rates and code generation performance for modified problems, it sometimes asked unnecessary questions for original problems that did not require clarification. Ideally, a perfect model should intelligently determine when to ask questions and when to refrain from doing so, but there is a gap between this ideal behavior and the performance of all evaluated models, including Okanagan.}

\JWRR{\textbf{Potential bias introduced by LLM-based evaluator.} This study employed an LLM-based evaluator to assess the quality of clarifying questions generated by the models and to generate answers to those questions. While this approach proved useful, it also introduced several limitations that potentially impacted the generalizability of the results:}
\begin{itemize}
\item \JWRR{Discrepancies existed between the Good Question Rate as determined by the automated evaluator and the manual evaluation performed by human annotators. The LLM-based evaluator tended to mark more questions as ``Good'' than the human annotators. This means that the evaluator might have overestimated the quality of clarifying questions in some cases. For example, it sometimes assigned ``Good'' question quality labels even when the models did not ask any clarifying questions. In terms of evaluation results, Good Question Rate can provide useful insights, but they may not always align with human judgment due to the potential bias of evaluator. } 
\item \JWRR{Good Question Rates are more easily affected by different prompt variants due to the evaluator. This aligns with our finding of discrepancies in Good Question Rate in RQ3. Thus, regarding the evaluation result, Good Question Rates are more vulnerable to prompt engineering, indicating the unreliability of this metric. } 
\item \JWRR{The False Recovery Rate for some models, such as CodeQwen1.5 Chat and DeepSeek Coder, was notably higher compared to Okanagan. This indicates that the evaluator sometimes provided relevant information to these models even when they did not ask any relevant clarifying questions. Consequently, this may have artificially inflated their Test Pass Rates and Pass@1 scores, complicating the assessment of their communication skills via these metrics.}
\end{itemize}

\begin{table*}
\centering
\begin{tabular}{p{6.7cm} p{6.5cm}} 
\hline
\textbf{Prompt Template} & \textbf{Output (Auto-Modified Problem)} \\
\hline

\begin{minipage}[t]{\linewidth}
Based on the knowledge that ambiguous problem descriptions can be created by introducing multiple valid interpretations or unspecified details, think step-by-step to rewrite the given coding problem description and make it ambiguous. Only output the modified problem description itself.\\

\textbf{Original Problem Description:}
\begin{lstlisting}[language=Python, basicstyle=\ttfamily\small, breaklines=true]
def truncate_number(number: float) -> float:
    """ Given a positive floating point number, it can be decomposed into
    an integer part (largest integer smaller than given number) and decimals
    (leftover part always smaller than 1).

    Return the decimal part of the number.
    >>> truncate_number(3.5)
    0.5
    """
\end{lstlisting}
\textbf{Ambiguous Problem:}
\end{minipage}
&

\begin{minipage}[t]{\linewidth}
\begin{lstlisting}[language=Python, basicstyle=\ttfamily\small, breaklines=true]
def truncate_number(number: float) -> float:
    """ Given a number, it can be decomposed into an integer part (largest integer smaller than given number) and decimals
    (leftover part always smaller than 1).

    Return the decimal part of the number.
    >>> truncate_number(3.5)
    0.5
    """
\end{lstlisting}
(Interpretation: ``a positive floating point number'' is changed to ``a number'', with the intention to make the problem ambiguous. Although this modification is made, it doesn't trigger clarifying questions before generating correct code, which is required for crafting modified problems in this study.)
\end{minipage}
\\
\hline

\begin{minipage}[t]{\linewidth}
Based on the knowledge that a problem becomes inconsistent if some statements in the description show conflict, think step-by-step to rewrite the given coding problem description and make it inconsistent. Only output the modified problem description itself.\\

\textbf{Original Problem Description:}
\begin{lstlisting}[language=Python, basicstyle=\ttfamily\small, breaklines=true]
from typing import List

def filter_by_substring(strings: List[str], substring: str) -> List[str]:
    """ Filter an input list of strings only for ones that contain given substring
    >>> filter_by_substring([], 'a')
    []
    >>> filter_by_substring(['abc', 'bacd', 'cde', 'array'], 'a')
    ['abc', 'bacd', 'array']
    """
\end{lstlisting}
\textbf{Inconsistent Problem:}
\end{minipage}
&

\begin{minipage}[t]{\linewidth}
\begin{lstlisting}[language=Python, basicstyle=\ttfamily\small, breaklines=true]
from typing import List

def filter_by_substring(strings: List[str], substring: str) -> List[str]:
    """ Filter an input list of strings only for ones that contain given substring
    >>> filter_by_substring([], 'a')
    []
    >>> filter_by_substring(['abc', 'bacd', 'cde', 'array'], 'a')
    ['abc', 'bacd', 'array']
    >>> filter_by_substring(['abc', 'bacd', 'cde', 'array'], 'b')
    ['abc', 'bacd']
    """
\end{lstlisting}
(Interpretation: a new test example is added to make the problem inconsistent, but this modification does not create any conflict or inconsistency.)
\end{minipage}
\\
\hline
\end{tabular}
\caption{Automatically modified problems using a LLM. Given a problem description from the HumanEval dataset, we used Google Gemini as the LLM to modify the problem description so that the modified problem is ambiguous, inconsistent, or incomplete.}
\label{tab:auto_modified_problems}
\end{table*}

\JWRR{\section{Implications and Future Work}}
\JWRR{Based on the results and analysis, we summarized the following implications as suggestions for future work:}

\begin{itemize}
\item \JWRR{\textbf{Develop better evaluation methods for communication ability in coding tasks.} Our main contribution is the new HumanEvalComm benchmark in evaluating LLMs and LLM agents on the communication aspect when the models are assigned coding tasks, but future research is needed to improve the evaluation. This includes 
\begin{enumerate}
        \item Exploring automated or semi-automated techniques to reduce reliance on human resources, making the method more accessible to resource-constrained environments.  
        \item Addressing biases in LLM-based evaluators, such as adopting approaches like the FM/LLM Jury~\cite{li2024software}.  
        \item Enhancing the accuracy and fairness of evaluation metrics to better measure and understand communication capabilities.  
        \item Creating diverse datasets with varying levels of clarification difficulty to more comprehensively assess the communication skills of models.  
    \end{enumerate}  
We believe this is an important angle in evaluation as more and more models in agentic flow and LLMs are starting to emerge for software engineering tasks in the near future.}
\item  \JWRR{\textbf{Move beyond the limitations of generative models.} In the evaluation, we found that for \textit{Incompleteness} category, some Code LLMs have extremely low results, potentially due to their generative nature that prefers to generate and complete code based on a statistical model, even when the description is obviously incomplete. Future work should go beyond the generative nature of LLMs to ``AI agent'' or ``AI assistants''~\cite{hassan2024rethinking} to further enhance communication capabilities.  }
\item  \JWRR{\textbf{Enhance reasoning capability and information-gathering capabilities. } In the evaluation, \textit{Inconsistency} category has the lowest communication rate among the three types. This indicates that more future work is needed to develop models with stronger reasoning capability to address the low performance in \textit{Inconsistency} category. Separately, from the RQ1 results in Table~\ref{Tab:sec5_3} and RQ4 results of different incomplete modifications in Section \ref{sec:diff_1p}, we noticed reduced Pass@1 and Test Pass Rates as the difficulty increases in the problem description. This potentially indicates a bottleneck in failing to fetch needed information for solving the coding tasks, particularly for more complex problem descriptions. Future work should focus on both model and evaluation setup to increase the model's ability to reason about the problem description and to get necessary information in challenging situations and real-world environments. } 
\item  \JWRR{\textbf{Improve the ability to determine when to stop asking questions. } Although the LLM agent approach, Okanagan, showed promising initial results in improving the metrics in the evaluation, one limitation of Okanagan is that it reduced the pass rates in the original HumanEval benchmark, due to asking unnecessary questions. Therefore, one future work in LLM agents is to address this shortcoming. This will potentially lead to much stronger communication capability as the model will know intelligently when to avoid redundant or irrelevant inquiries.  }
\end{itemize}

\section{Threats to Validity}
\label{sec7}
\textbf{Construct validity.} This threat relates to the potential incorrectness in manual modifications of problems in HumanEvalComm. To mitigate this threat, we have manually checked and verified all of the problems more than three times, and each time they discussed the problems they didn't reach a consensus. Although we have tried our best efforts, there still may be some corner cases where \JWR{1) the modified problems do not match the definition of \textit{Ambiguity} and \textit{Inconsistency}, and 2) the modified problems suffer from subjective variations and are not consistent with other problems.}

\textbf{Internal validity.} This threat relates to the internal parameters such as the parameters in open-source Code LLMs and ChatGPT that could potentially affect the results. To mitigate this threat, we use most of the default parameters when running open-source Code LLMs and ChatGPT. For open-source models, we set \textit{max\_new\_tokens} as 512 to save computing resources and used default values for other parameters. For ChatGPT, we used temperature as 1.0 and $n$ as 1 in the OpenAI API. \JIEW{Another threat relates to the effectiveness of the LLM-based evaluator used in the evaluation. As mentioned previously, to mitigate this issue, we have optimized the prompt for the LLM-based evaluator several times and checked the results manually. Besides, the LLM-based evaluator is used equally for all models in the evaluation, so this threat does not affect the relative ranking of the results for all models. }

\textbf{External validity.} This relates to the generality of the communication capability of the models on other benchmarks. To mitigate this issue, we extensively report and analyze the results with statistical testing that reports p-value. To reduce the risk introduced by randomness in our investigation, we also added metrics such as mean, variance, max diff in the results. Thus, these results can be potentially adapted for other datasets. However, since we have not tested this, we cannot make a sound claim regarding the communication capability of the models on another dataset. Another threat is related to the implementation of evaluated models. We directly call OpenAI API to get ChatGPT results. We implemented Okanagan in Python that calls OpenAI API. For CodeLlama and other open-source models, we downloaded the model from HuggingFace and perform model inference on UBC ARC Sockeye. From the evaluation results, we believe that our implementation reflects the original methods. \JWR{Another threat concerns the generality and reliability of the metrics we used in our evaluations. To address this, we have added RQ3 to conduct manual evaluation and use its result to compare with the proposed metrics. We have also discussed the potential impact of LLM-based evaluator on the reliability of the metrics. } To ensure the reproducibility of the evaluation results, we report the result of the case study extensively and release our complete code and dataset. This can allow other researchers to reproduce and extend our experiments in the case study.


\section{Related Work}
\label{sec8}
\textbf{Code Generation with Large Language Models.} In recent years, the field of code generation has seen a significant shift with the large language models. For example, Codex~\cite{chen2021evaluating}, fine-tuned on GPT-3~\cite{brown2020language} on a large corpus of source code data, is capable of generating code for 47/164 problems in the HumanEval dataset in a single run, a benchmark for code generation task. Codex became the core model for the Copilot~\cite{ziegler2022productivity}, an AI-powered coding assistant developed by GitHub. After Codex, a couple of models similar to Codex but with smaller size were then developed, including GPT-J~\cite{wang2021gpt}, CodeParrot~\cite{huggingface-codeparrot}, PolyCoder~\cite{xu2022systematic}. AlphaCode~\cite{li2022competition}, with size comparable to Codex, was trained on Github data and fine-tuned on competition-level programming problems. It exceeded half of the competitors in coding competitions of CodeForces, a well-known online competitive programming platform. CodeGen~\cite{nijkamp2022codegen} was trained on both natural language and programming language data for code generations with multi-turn prompts. \JIE{Recently, newer models such as CodeLlama~\cite{roziere2023code}, DeepSeek Coder~\cite{guo2024deepseek} and CodeQwen1.5 Chat~\cite{bai2023qwen} continued to achieve higher performance in benchmark such as HumanEval.} However, the level of communication skills of these models is not emphasized and evaluated. These models are evaluated by generating code in one or multiple attempts from one-off problem descriptions, without further information from conversations. Therefore, when the input problem description is error-prone or incomplete, the model still has to generate the code without the chance to clarify critical questions. Our work serves as an exploration to address this usability problem.
    
\textbf{Self-Correct LLMs and LLM Agent in AI.} Recently, a promising approach to improve the output efficiency of large language models is \textit{self-correction}~\cite{pan2023automatically}. In the self-correction approach, the LLM uses the feedback guided or prompted by itself to refine its results. One popular category of work uses human feedback to refine their results directly~\cite{kreutzer2018, glaese2022, ouyang2022, scheurer2023, fernandes2023}. Other studies employed different strategies to self-correct LLMs using automated feedback such as self-training \cite{huang2022, bai2022b}, generate-then-rank \cite{he2023, weng2023}, feedback-guided decoding \cite{yang2022a, xie2023}, iterative post-hoc revision \cite{zhang2023a, jiang2023}, etc. Furthermore, the advances in LLM have also brought much progress in LLM-based agents, with different modules including Planning, Memory, Profile, and Action~\cite{xi2023rise,wang2023survey}, and various agent categories, such as Tool Agent, Simulation Agent, Web Agent, Game Agent, etc. Our work also includes the evaluation of the LLM agent approach, Okanagan, which has an additional round with reflection as the thinking pattern.

\textbf{LLM Agent for Code Generation.} Although still at an early stage, recently, there has been a rising stream of research efforts to employ \JWR{LLM agents for the task of code generation~\cite{yang2024swe,zhang2024codeagent,auto-gpt,yao2023react,cognitionlabs,shinn2023reflexion,schick2023toolformer}. In the AI community, the ReAct paper proposes a method for integrating reasoning and acting in language models to enhance their decision-making capabilities~\cite{yao2023react}. Reflexion introduces a framework where language agents improve their performance through verbal reinforcement learning, simulating human-like reflection and adaptation~\cite{shinn2023reflexion}. Toolformer demonstrates that language models can autonomously learn to use external tools, such as search engines and calculators, to improve task performance~\cite{schick2023toolformer}. SWE-Agent proposes a new agent designed specifically for software engineering tasks, leveraging large language models to enhance productivity and decision-making~\cite{yang2024swe}. CoderAgent introduces a model aimed at automating coding tasks by utilizing advanced language models to generate and refine code snippets efficiently~\cite{zhang2024codeagent}.}

In the field of software engineering, RepairAgent~\cite{bouzenia2024repairagent} is one of the first to use an LLM-based agent for program repair and code generation. This work follows the previous work in augmenting LLMs with API tools. Recently, CoRE \cite{xu2024core} has been proposed as a system that enables agent programming by using LLM as interpreters to process and execute natural language instructions. Following a similar spirit of LLM agent, TICODER~\cite{fakhoury2024llm} is proposed as a test-driven interactive workflow for more accurate code generation. Similarly, De-Hallucinator \cite{eghbali2024hallucinator} is proposed as a code completion method that combines retrieval-based code generation and iterative querying of the model.    
\JIEW{Different from the above works, the proposed Okanagan in our work focuses on enhancing the communication capabilities of LLM for code generation tasks. To the best of our knowledge, we are the first to study and compare the communication capabilities of LLM agent and Code LLMs in code generation tasks.}

\section{Conclusions}
\label{sec9}
In this paper, we showed an initial step in the empirical study of the communication skills of LLMs in evaluating code clarification and code generation. We argue that the proficiency of communication skills of LLMs is necessary for AI systems to generate code with high standards, and, in the long term, to ask questions to acquire information that is just enough to complete their tasks. We believe that elevated communication skills should be viewed as an important factor in bridging the gap between LLMs and top-notch software developers. Although it needs additional conversational inputs, we believe it is still necessary and worthwhile to evaluate this communication capability for coding tasks. 

\JIE{As a first step toward this effort, we created HumanEvalComm to evaluate the degree of communication skills. Based on the new benchmark, we comprehensively evaluated different Code LLMs with the communication lens, where certain information is manually modified in the original problem description. Furthermore, we proposed an LLM-based agent approach, Okanagan, to identify and ask questions in ambiguous parts of code and descriptions for further refining the generated code. We found that modifying the problem description greatly reduced Test Pass Rates and Pass@1 with statistical significance. In terms of communication skills, more than 60\% of responses from Code LLMs still generate code rather than ask questions when the problem descriptions are manually modified. We also find that, compared with LLM such as ChatGPT 3.5, Okanagan, as a LLM agent approach, can effectively increase Communication Rate and Good Question Rate, and thus boost Test Pass Rate and Pass@1 when the problem descriptions are modified based on a clarification type.}


Besides benchmarks, techniques to further improve the communication skills of LLMs can be the next steps in future work. Another interesting angle is to study how to tune the model to switch between under-communicating, effective-communicating, and over-communicating. We envision that different AI programming agents in the future will have various levels and styles of communication ability. This work can be seen as the first step toward evaluating the communication skills of Code LLMs and LLM agents. Our benchmark and replication package are made public at \href{https://github.com/jie-jw-wu/human-eval-comm}{https://github.com/jie-jw-wu/human-eval-comm}.

\JWRR{\section{Acknowledgments}}
\JWRR{We would like to thank the anonymous reviewers for their valuable feedback and suggestions, which helped improve the quality of this work significantly. We also extend our gratitude to six graduate students at the University of British Columbia who conducted the manual evaluation. Finally, we would like to thank the authors of the Non-Determinism research of ChatGPT~\cite{ouyang2023llm} and IdentityChain~\cite{min2023beyond} for releasing their code. This research is supported by a grant from Natural Sciences and Engineering Research Council of Canada RGPIN-2019-05175.}
 
\section*{APPENDICES}
\appendix

\section{APPENDIX: Prompt Templates In HumanEvalComm}

\begin{table}[H]
\centering
\begin{tabular}{p{10cm}p{3cm}}
\hline
\textbf{Prompt Template} & \textbf{Description} \\
\hline
\texttt{You are an expert software developer who writes high quality code. With below information, please either generate Python3 code (Respond directly with code only with markdown), or ask clarifying questions: \{problem\}} & Initial-round prompt for the model. Used in evaluation. \\
\hline
\texttt{The original description of a coding problem is modified so that the requirements become inconsistent, incomplete, or ambiguous. Given the modified description, some clarifying questions were raised to clarify the description. Given the original and modified problem description, evaluate the quality of the clarifying questions. Please provide an integer representing the quality of questions (3: Good questions that recover the modified requirements; 2: Fair questions but they cannot help recover the modified requirements; 1: No questions). \newline QUALITY=[your int] \newline Please also provide answers to the clarifying questions to recover the modified requirements in the original problem description compared to the modified one. If there is no clarifying questions at all, return empty answers. \newline ANSWERS=\texttt{[your answer]} \newline Please strictly follow the format QUALITY=[the int] and ANSWERS=\texttt{[the answer]} in the response! Surround your answer with markdown! \newline \newline \#\#\# Questions: \{clarifying\_questions\} \newline \#\#\# Modified Problem Description: \{problem\} \newline \#\#\# Original Description: \{missing\_information\} \newline} & Prompt for the LLM-based evaluator. Used in evaluation. \\
\hline
\texttt{\{problem\} \newline \{clarifying\_questions\} \newline \{answer\_to\_questions\}\newline
Given above conversations, generate Python code directly (Markdown) to solve the coding problem: } & Second-round prompt for the model. Used in evaluation. \\
\hline
\end{tabular}
\caption{\JWR{Detailed Prompt Templates used in the HumanEvalComm evaluation in this research. We tested different prompt templates and checked their impact to the results. }}
\label{tab:apx_prompts_comm} 
\end{table} 

\begin{table*}
\centering
\begin{center} 
\begin{tabular}{p{1.5cm}p{11cm}}
\toprule
1st Prompt Variant &
You are an expert software developer specialized in Python3. Review the problem description and: \newline
- If the description provides enough detail, proceed by writing the required Python3 code using markdown. \newline
\textbf{OR} \newline
- If further details are needed to ensure the solution is accurate and complete, please list any specific questions you have about the problem. \newline
\textbf{\#\#\# Problem Description:} \newline
\{problem\} \\ \midrule
1st Prompt One-Shot &
You are an expert software developer specialized in Python3. In each turn, review the problem description and: \newline
- If the description provides accurate detail, proceed by writing the required Python3 code using markdown. \newline
\textbf{OR} \newline
- If further details are needed to ensure the solution is accurate and complete, list clarifying questions about the problem without writing code. \newline
\textbf{\#\#\# Problem Description:} \newline
\begin{verbatim}
def candidate(x: int, y: int):
    """
    Return 2^n modulo p (be aware of numerics).
    >>> candidate(3, 5)
    3
    >>> candidate(1101, 101)
    2
    >>> candidate(0, 101)
    1
    >>> candidate(3, 11)
    8
    >>> candidate(100, 101)
    1
    """
\end{verbatim} \newline
\textbf{\#\#\# Response:} \newline
The problem description seems to be incomplete. It provides a function signature and some test cases, but it doesn\texttt{'}t specify what the function is supposed to do. The docstring suggests that the function should return \(2^n\) modulo \(p\), but the function parameters (x, y) don\texttt{'}t match with the variables in the docstring (n, p). \newline
Here are my questions to clarify the problem: \newline
1. What are the roles of the parameters x and y in the function? \newline
2. Should the function return \(2^n\) modulo \(p\) as suggested in the docstring? If so, which parameter corresponds to n and which corresponds to p? \newline
3. The test cases provided in the docstring don\texttt{'}t seem to match the function description. Could you please provide correct test cases? \newline 
\textbf{\#\#\# Problem Description:} \newline
\{problem\}\newline
\textbf{\#\#\# Response:} \\
\bottomrule
\end{tabular}
\end{center}
\caption{\JWR{Prompt variants used in prompt engineering for RQ4.}}
\label{Tab:apx_prompt_variants}
\end{table*}

\begin{table}[H]
\centering
\begin{tabular}{p{10cm}p{3cm}}
\hline
\textbf{Prompt Template} & \textbf{Description} \\
\hline
\texttt{Generate Python code directly (Markdown) to solve the coding problem. \newline \newline \{problem\}} & Prompt to generate code in Okanagan. \\
\hline
\texttt{Given the programming problem, ask clarifying questions if the requirements in the given problem description are incomplete, inconsistent or ambiguous for solving the problem correctly and passing the tests. \newline If no need to ask clarifying questions, return strictly 'NO\_QUESTIONS' only. Otherwise, return the clarifying questions. \newline \newline \#\#\# Problem:\newline \{problem\}} & Prompt to ask questions in Okanagan. \\
\hline
\end{tabular}
\caption{\JWR{Detailed Prompt Templates used in Okanagan as an LLM-based agent approach proposed in this research. }}
\label{tab:apx_prompts_ok}
\end{table} 

\bibliographystyle{ACM-Reference-Format}
\bibliography{references}

\end{document}